\documentclass[11pt]{llncs}
\usepackage{llncsdoc}
\RequirePackage{Settings}
\RequirePackage{Makros}
\begin{document}



\title{%
Formal Derivation of Concurrent Garbage Collectors
} 

\author{Dusko Pavlovic\inst{1}
  \and Peter Pepper\inst{2}
  \and Douglas R. Smith\inst{1}}

\institute{Kestrel Institute, 
  Palo Alto, California \\ 
  \email{\{dusko,smith\}\char64kestrel.edu} 
  \and
  Technische Universit\"at Berlin and Fraunhofer FIRST, Berlin\\
\email{pepper\char64cs.tu-berlin.de}}

\maketitle

\begin{abstract}
  Concurrent garbage collectors are notoriously difficult to implement
  correctly.  Previous approaches to the issue of producing correct
  collectors have mainly been based on posit-and-prove verification or
  on the application of domain-specific templates and transformations.
  We show how to derive the upper reaches of a family of concurrent
  garbage collectors by refinement from a formal specification,
  emphasizing the application of domain-independent design theories
  and transformations.  A key contribution is an extension to the
  classical lattice-theoretic fixpoint theorems to account for the
  dynamics of concurrent mutation and collection.

\end{abstract}



\section{Introduction}
\label{sec:Introduction}

\emph{Concurrent collectors are extremely complex and error-prone.
  Since such collectors now form part of of the trusted computing base
  of a large portion of the world's mission-critical software
  infrastructure, such unreliability is unacceptable}
\cite{Vechev+2006}. Therefore it is a worthwhile if not mandatory
endeavor to provide means by which the quality of such software can be
improved -- without doing harm to the productivity of the programmers.

The latter aspect still is a major obstacle in verification-oriented
systems. Interactive theorem provers may need thousands of lines of
proof scripts or hundreds of lemmas in order to cope with serious
collectors (see e.g.
\cite{McCreight+2007,Russinoff:1994,Doligez+1994}). But also fully
automated verifiers exhibit problems. As can be seen e.g. in
\cite{HawPet:2009} even the verification of a simplified collector
necessitates such a large amount of complex properties that the
specification may easily become faulty itself.

These considerations show a first mandatory prerequisite for the
development of correct software of realistic size and complexity: not
only the software but also its correctness proof need to be
\emph{modularized}. However, such a modularization is not enough. Even
when it has been successfully verified that all requested properties
are fulfilled by the software, it remains open, whether these
properties taken together do indeed specify the intended
behavior. This is an external judgment that lies outside of any
verification system. Evidently, such judgments are easier and more
trustworthy, when the properties are few, simple and easy to grasp.

Finally there is a third aspect which needs to be addressed by a
development methodology. Garbage collectors -- like most software
products -- come in a plethora of possible variations, each addressing
specific quality or efficiency goals. When each of these variations is
verified separately, a tremendous duplication of work is generated. On
the other hand it is extremely difficult to analyze for a slightly
modified algorithm, which properties and proofs can remain unchanged,
which are superfluous and which need to be added or redone.

We propose here a development method, which addresses the
aforementioned issues and which has already been successfully applied
to complex problems, for example real-world-size planning and
scheduling tasks {\cite{SmithD9605,SmithD9606}}. The method bases on
the concept of \emph{specification refinement}. Two major aspects of
this concept are illustrated in \Figs{fig:Pushout}
and~\ref{fig:Refinement}.

\subsection{Sequential vs. Concurrent Garbage Collection}
\label{sec:Sequential-concurrent}

The very first garbage collectors, which essentially go back to
McCarthy's original design \cite{McCarthy:1960}, were 
\emph{stop-the-world} collectors. That is, the Mutator was completely
laid to sleep, while the Collector did its recycling. This approach
leads to potentially very long pauses, which are nowadays considered
to be unacceptable.

The idea of having the Collector run concurrently with the Mutator
goes back to the seminal papers of Dijkstra et
al.~\cite{Dijkstra+1978} and Steele \cite{Steele:1975} (which were
followed by many other papers trying to improve the algorithm or its
verification). The Doligez-Leroy-Gonthier algorithm (short: DLG) that
was developed for the Concurrent CAML Light system
\cite{Doligez+1994,Doligez+1993}, is considered an important
milestone, since it not only takes many practical complications of
real-world collectors into account, but also generalizes from a single
Mutator to many Mutators.

The transition to concurrent garbage collection necessitates a
\emph{trade-off between the precision of the Collector and the degree
of concurrency it provides} \cite{Vechev+2006}: the higher the degree
of concurrency, the more garbage nodes will be overlooked. However,
this is no major concern in practice, since the escaped garbage nodes
will be found in the next collection cycle.

\subsection{Abstract and Concrete Problems}
\label{sec:Problems}

\begin{figure}[htbp]
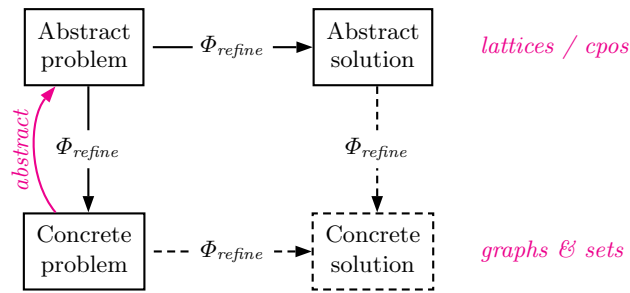

\begin{center}
\newcommand\BoxIt[2][solid]{%
\psframebox[linestyle=#1]{\begin{tabular}{c}#2\end{tabular}}
}
\begin{tabular}{c@{\hspace{2.1cm}}c@{\hspace{.5cm}}l}
\rnode{AP}{\BoxIt{Abstract\\problem}} & \rnode{AS}{\BoxIt{Abstract\\solution}}
& {\magenta\footnotesize\emph{lattices / cpos}}
\\[2cm]
\rnode{CP}{\BoxIt{Concrete\\problem}} & \rnode{CS}{\BoxIt[dashed]{Concrete\\solution}}
& {\magenta\footnotesize\emph{graphs \& sets}}
\end{tabular}
\nccurve[linecolor=magenta,angleA=130,angleB=230]{->}{CP}{AP}
\Aput{\rput{90}(0,0){\footnotesize\magenta\emph{abstract}}}
{%
\ncline{->}{AP}{AS}\mput*{$\Phi_{\mathit{refine}}$}
\ncline{->}{AP}{CP}\mput*{$\Phi_{\mathit{refine}}$}
\psset{linestyle=dashed}
\ncline{->}{AS}{CS}\mput*{$\Phi_{\mathit{refine}}$}
\ncline{->}{CP}{CS}\mput*{$\Phi_{\mathit{refine}}$}
}%
\caption{Abstract and concrete problems and their solutions}
\label{fig:Pushout}
\end{center}
\end{figure}

\Fig{fig:Pushout} describes the way in which we come from abstract
problems to concrete solutions. (1) Suppose we have an \emph{abstract
  problem description}, that is, a collection of types, operations and
properties that together describe a certain problem. (2) For this
abstract problem we then develop an \emph{abstract solution}, that is,
an abstract implementation that fulfills all the requested properties.
(3) When we now have a \emph{concrete problem} that is an instance of
our abstract problem (since it meets all its properties), then we can
(4) automatically derive a \emph{concrete solution} by instantiating
the abstract solution correspondingly. Ideally the abstract
problem/solution pairs can even be found in a library like the one of
the Specware system \cite{Specware03}.

For example, in the subsequent sections of this paper we will consider
the abstract problem of finding fixed points in lattices or cpos and
several solutions for this problem. Then we will show that garbage
collection is an instance of this abstract problem by considering the
concrete graphs and sets as instances of the more abstract
lattices. This way our abstract solutions carry over to concrete
solutions for the garbage collection problem.

Technically all our problem and solution descriptions are algebraic
and coalgebraic specifications (as will be defined more precisely
later), which are usually underspecified and thus possess many models.
``Solutions'' are treated as borderline cases of such specifications,
which are directly translatable into code of some given programming
language. (This concept has nowadays been popularized as ``automatic
code generation from models''.) The formal connections between the
various specifications are given by certain kinds of refinement
morphisms, and the derivation of the concrete solution from the other
parts is formally a pushout construction from category
theory\footnote{A morphism $\Phi$ from specification $S$ to
  specification $T$ is given by a type-consistent mapping of the type,
  function, and predicate symbols of $S$ to derived types, functions,
  and predicates in $T$.  The mapping is a specification morphism if
  the axioms of $S$ translate to theorems of $T$.  A pushout
  construction is used to compose specifications.  More detail on the
  category of specifications may be found in
  \cite{SmithD9909,Specware03,PPS:2003}}.

\Fig{fig:Pushout} also illustrates another aspect of our
methodological way of proceeding. When we are confronted with a
concrete problem, we try to extract from it a more abstract problem
that represents the core of the given task. Even though this looks
like additional effort at first sight, it is usually a worthwhile
endeavor. First of all, we obtain the desired modularization of the
derivation and verification. Secondly, the concentration on the kernel
of the problem usually simplifies the finding of the (abstract)
solutions. And last but not least we can often come up with variations
on the theme that would have been buried under the bulk of details
otherwise. As is pointed out in \Fig{fig:Pushout} the introduction of
the details of the concrete problem can be done almost automatically
and thus does not really cause additional work.

This principle of working with an abstract view of the concrete
problem can also be found in other approaches, for example in
\cite{McCreight+2007,HawPet:2009}. But there the principle is more
implicitly used (in statements such as \emph{Correctness means that
each of these procedures faithfully represent the abstract state}
\cite{HawPet:2009}), whereas we make the abstraction/concretization
into an explicitly available development tool, based on a rigorous
notion of morphisms.

\subsection{Development by Refinement}
\label{sec:Development-Refinement}

\begin{figure}[htbp]
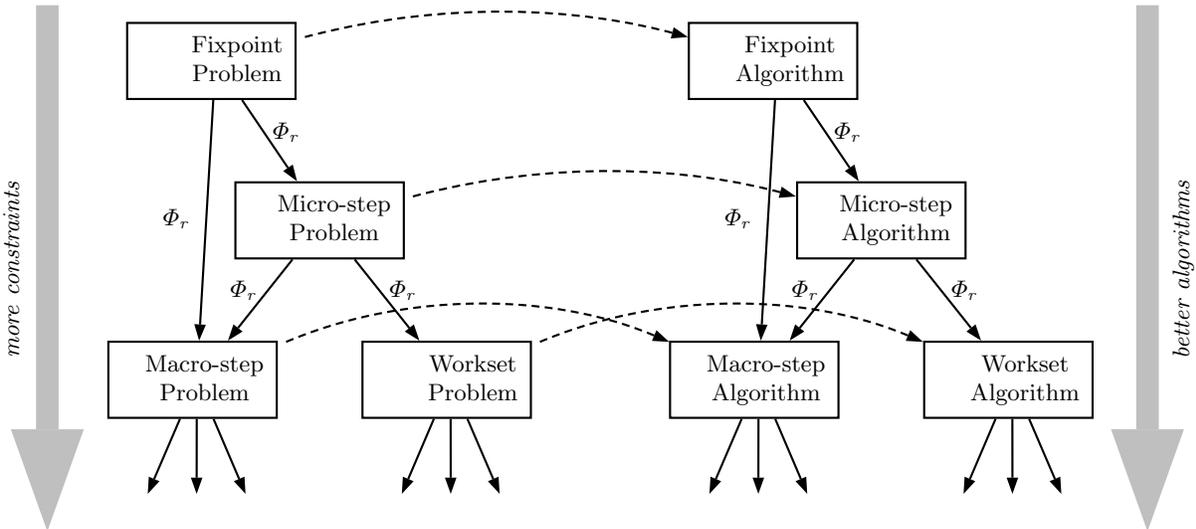

\begin{center}
\Unit{1mm}
\newcommand\BoxIt[2][solid]{%
\psframebox[linestyle=#1]{\parbox{2.0cm}{\hfill\begin{tabular}{c}#2\end{tabular}\hfill}}
}
\newcommand\RefHom{$\Phi_{\mathit{r}}$}
\pcline[linecolor=lightgray,linewidth=3mm]{->}(-3,35)(-3,-35)%
\Bput{\rput{90}{\emph{more constraints}}}
\begin{tabular}{@{}c@{\hspace{.5cm}}c@{\hspace{.5cm}}c@{\hspace{.5cm}}c@{}}
\mc3{c}{\rnode{FP}{\BoxIt{Fixpoint\\Problem}}} &
\\[40pt]
&  \mc3{c}{\rnode{MiP}{\BoxIt{Micro-step\\Problem}}} 
\\[40pt]
& \rnode{MaP}{\BoxIt{Macro-step\\Problem}} & & \rnode{WP}{\BoxIt{Workset\\Problem}}
\\[30pt]
& \pnode(-2em,0){AP1}\pnode(0em,0){AP2}\pnode(2em,0){AP3} &
& \pnode(-2em,0){BP1}\pnode(0em,0){BP2}\pnode(2em,0){BP3}
\end{tabular}
\hspace{1cm}
\begin{tabular}{@{}c@{\hspace{.5cm}}c@{\hspace{.5cm}}c@{\hspace{.5cm}}c@{}}
\mc3{c}{\rnode{FS}{\BoxIt{Fixpoint\\Algorithm}}} &
\\[40pt]
&  \mc3{c}{\rnode{MiS}{\BoxIt{Micro-step\\Algorithm}}} 
\\[40pt]
& \rnode{MaS}{\BoxIt{Macro-step\\Algorithm}} & & \rnode{WS}{\BoxIt{Workset\\Algorithm}}
\\[30pt]
& \pnode(-2em,0){AS1}\pnode(0em,0){AS2}\pnode(2em,0){AS3} &
& \pnode(-2em,0){BS1}\pnode(0em,0){BS2}\pnode(2em,0){BS3}
\end{tabular}
\pcline[linecolor=lightgray,linewidth=3mm]{->}(5,35)(5,-35)
\Aput{\rput{90}{\emph{better algorithms}}}
\ncline{->}{FP}{MaP}\Bput{\RefHom}
\ncline{->}{FP}{MiP}\Aput[0pt]{\RefHom}
\ncline{->}{MaP}{AP1}
\ncline{->}{MaP}{AP2}
\ncline{->}{MaP}{AP3}
\ncline{->}{MiP}{WP}\Aput[0pt]{\RefHom}
\ncline{->}{MiP}{MaP}\Bput[0pt]{\RefHom}
\ncline{->}{WP}{BP1}
\ncline{->}{WP}{BP2}
\ncline{->}{WP}{BP3}
\ncline{->}{FS}{MaS}\Bput{\RefHom}
\ncline{->}{FS}{MiS}\Aput[0pt]{\RefHom}
\ncline{->}{MaS}{AS1}
\ncline{->}{MaS}{AS2}
\ncline{->}{MaS}{AS3}
\ncline{->}{MiS}{WS}\Aput[0pt]{\RefHom}
\ncline{->}{MiS}{MaS}\Bput[0pt]{\RefHom}
\ncline{->}{WS}{BS1}
\ncline{->}{WS}{BS2}
\ncline{->}{WS}{BS3}
\nccurve[linestyle=dashed,angleA=15,angleB=165]{->}{FP}{FS}
\nccurve[linestyle=dashed,angleA=15,angleB=165]{->}{MiP}{MiS}
\nccurve[linestyle=dashed,angleA=23,angleB=157]{->}{MaP}{MaS}
\nccurve[linestyle=dashed,angleA=23,angleB=157]{->}{WP}{WS}
\caption{Refinement of problems (and solutions)}
\label{fig:Refinement}
\end{center}
\end{figure}

\Fig{fig:Refinement} illustrates the second essential aspect of our
method. We do not work with a single problem/solution pair and their
concrete instances. Rather we construct a whole ``family tree'' (which
actually may be a dag) of more and more refined problems, each giving
rise to more and more refined solutions. On the problem side
``refined'' essentially means that we have additional properties, on
the solution side ``refined'' essentially means that we have better
algorithms, e.g. more efficient, more robust, more concurrent etc.

This way of proceeding has the primary advantage that it allows us to
reuse verification and development efforts. Suppose that at some point
in the tree we want to design a new variation. This is reflected in a
new refinement child of the current specification, to which certain
properties are added. In the accordingly modified new solution we need
only prove those properties that have been added; everything else is
inherited.

More details about the method sketched above can be found in
\cite{SmithD9909}. The remainder of this paper will make things more
precise by presenting concrete examples.

In an earlier paper \cite{PPS:2003} we have presented one exemplary
development of a garbage collector from an initial non-executable
specification to an executable implementation. But -- as was
critically noted in \cite{Vechev+2006} -- we did ``not explore an
algorithm space''. Such an exploration is the main purpose of the
present paper. This is a similar goal as that of Vechev et
al.~\cite{Vechev+2006}: they start from a generic algorithm, which is
parameterized by an underspecified function, such that different
instantiations of this function lead to different collection
algorithms. A primary concern of \cite{Vechev+2006} is the possibility
to combine various ``design dimensions'' in a very flexible way. By
contrast to their approach we study the family tree of specifications
and implementations that can be systematically derived using formal
refinements. (The interchangeability of some of these refinements
actually makes the family tree into a family dag.) So our focus is on
the \emph{method} of refinement and its potential tool support and not
on garbage collection as such. Moreover, whereas both our earlier
paper \cite{PPS:2003} and the work of Vechev et al.~\cite{Vechev+2006}
mostly concentrate on one phase of garbage collection -- namely the
marking phase -- the present paper addresses the whole task of garbage
collection. In addition, we do not only consider mark-and-sweep
collectors but also copying collectors. Last but not least, we base
the whole treatment of garbage collection on very fundamental
mathematical principles, namely lattices and fixed points.

\subsection{Data Reification}
\label{sec:Reification}

The final efficiency of most practical garbage collection algorithms
depends on the use of clever data representations. Standard techniques
range from the classical stacks or queues to bit maps, overlayed
pointers, so-called dirty bits, color toggling and so forth.

In our approach all these designs fall under the paradigm of data
reification morphisms. This means that we can work throughout our
developments with high-level abstract data structures such as graphs
and sets in order to specify and verify the algorithmic aspects in the
clearest possible way. It will be only at the end of the derivation
that the high-level data structures are implemented by concrete data
structures, which are chosen based on their efficiency in the given
context. This step is widely automatic in systems like Specware
{\cite{Specware03}}, including many low-level optimizations. Since this
is very technical and can be done almost automatically by advanced
systems, we will only touch this part very briefly and sketchy here.

\subsection{Summary of Results}
\label{sec:Results}

We present a methodology that allows us to derive a wide variety of
garbage collection algorithms in a systematic way. This approach not
only modularizes the resulting programs but also the derivation
process itself such that the verification is split into small and
easy-to-comprehend pieces, allowing considerable reuse of proofs. In
more detail, we present the following results:

\begin{itemize}
\item We start by presenting a ``dynamic'' generalization of the
well-known fixed-point results of lattice theory.

\item This basis is presented as a general specification that covers a
whole range of implementations. We call this the ``micro-step''
approach.

\item These fixpoint-based specifications can be refined further to
more and more detailed designs, which correspond to the major
algorithms found in today's literature.
\end{itemize}

Even though we cannot present all the algorithms in full detail, we
can at least show ``in principle'', how a whole variety of important
and practical algorithms come out from our refinement process. These
include above all the (DLG) algorithm of Doligez, Leroy and Gonthier
\cite{Doligez+1994} -- which sometimes is considered the culmination
of concurrent collector development \cite{Azatchi+2003} -- and its
descendants.




\section{Notes on Garbage Collection}
\label{sec:Notes}

Even though our approach starts from very abstract and high-level
mathematical concepts -- viz. lattices and fixed points
(\Sec{sec:Fixpoints}) -- and takes several refinement steps
(\Sec{sec:Refinements}) before it ends with some special aspects of
garbage collection (\Sec{sec:Roots}), it is helpful to motivate our
main design decisions by having the concrete application of garbage
collection in mind.
Actually, we perceive three major stages of refinements:

\begin{enumerate}
\item We start from a ``purely mathematical problem'', namely lattices
  (actually cpos) and fixed points. On this level we derive the core
  properties that mark off the solution space.
\item Then we proceed to ``abstract garbage collection''; that is, we
  model the problem by graphs and sets. This intermediate stage can on
  the one hand be easily shown to be an instance of the lattice-based
  abstraction; but on the other hand it already refers to important
  aspects of the concrete garbage collection problem. Hence, all
  algorithmically relevant aspects can be dealt with on this level.
\item The final step introduces the various specialized data
  structures, write barriers and the like that go to make a realistic
  garbage collector. (This final step will only be roughly sketched in
  this paper.)
\end{enumerate}

\subsection{Architecture and Basic Terminology}
\label{sec:Architecture}

Before we delve into the formal derivation we want to clarify the
basic setting and the terminology that we use here. This is best done
at the intermediate level of abstraction, where the garbage collection
problem is formulated in terms of graphs and sets.

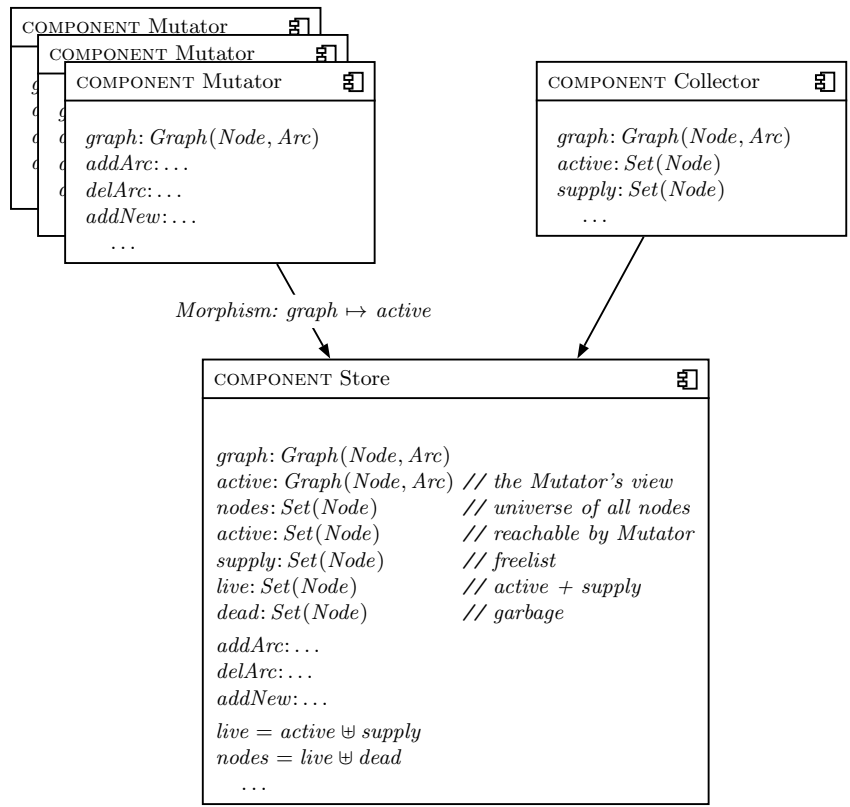
\begin{figure}[t]   
\begin{center}
\Unit{1mm}
{\small
\begin{spec}[plain,box=\progA,width=40\unit]
graph: Graph(Node,Arc)
addArc: ...
delArc: ...
addNew: ...
\quad\dots
\end{spec}
}
\def\CompA{
\SetComponentWidth{46}
\SetComponentHeight{30}
\Component Mutator={Mutator}{\small\usebox\progA}
}
{\small
\begin{spec}[plain,box=\progB,width=40\unit]
graph: Graph(Node,Arc)
active: Set(Node)
supply: Set(Node)
\quad\dots
\end{spec}
}
\def\CompB{
\SetComponentWidth{46}
\SetComponentHeight{26}
\Component Collector={Collector}{\small\usebox\progB}
}
{\small
\begin{spec}[plain,box=\progC,width=71\unit]
graph: Graph(Node,Arc)
active: Graph(Node,Arc)	\- the Mutator's view
nodes: Set(Node)		\- universe of all nodes
active: Set(Node)		\- reachable by Mutator
supply: Set(Node)	 \- freelist
live: Set(Node)		\- active + supply
dead: Set(Node)		\- garbage

addArc: \dots
delArc: \dots
addNew: \dots

live = active \uplus supply
nodes = live \uplus dead
\quad\dots
\end{spec}
}
\def\CompC{
\SetComponentWidth{75}
\SetComponentHeight{66}
\Component Store={Store}{\small\usebox\progC}
}
\psscalebox{.9}{%
\begin{pspicture}(0,0)(120,118)

\rput[tl](-8,118){\rnode{CompA}{\CompA}}
\rput[tl](-4,114){\rnode{CompA}{\CompA}}
\rput[tl](0,110){\rnode{CompA}{\CompA}}

\rput[tr](120,110){\rnode{CompB}{\CompB}}

\rput[b](60,00){\rnode{CompC}{\CompC}}

\ncline{->}{CompA}{CompC}
\mput*{\footnotesize\emph{Morphism: graph $\mapsto$ active}}

\ncline{->}{CompB}{CompC}
\end{pspicture}%
}
\caption{The system architecture}
\label{fig:Architecture}
\end{center}
\end{figure}

We modularize the problem by way of three kinds of components (using
a UML-inspired representation; see \Fig{fig:Architecture}). The
\(Mutators\) represent the activities of all programs that use the
heap. These activities base on primitive operations that are provided
by the component \(Store\), which represents the memory management
system (as part of the runtime system or operating system). Finally
the task of the garbage collection is performed by a component
\(Collector\).

The \(Mutator\) operates on a \(graph\), which is a data structure of
type \(Graph\). It can essentially perform three primitive
operations:%
\footnote{%
This considerably simplifies the memory model
used in the famous DLG algorithm \cite{Doligez+1994,Doligez+1993},
where the Mutator has eight operations. However, the essence of these
operations is captured by our three operations above. (We will come
back to this issue in \Sec{sec:Roots}.)  }%

\begin{itemize}
\item \strut\(addArc(a,b)\): add a new arc between two nodes \(a\) and
\(b\).
\item \strut\(delArc(a,b)\): delete the arc between \(a\) and \(b\).
  This may have the effect that \(b\) and other nodes reachable from
  \(b\) become unreachable (``garbage'').
\item \strut\(addNew(a)\): allocate a new node \(b\) (from the
freelist) and attach it by an arc from \(a\). This reflects the fact
that in reality \emph{alloc} operations return a pointer, which is
stored in some field (variable, register, heap cell) of the
Mutator. Hence, the new node is immediately linked to the Mutator's
graph.
\end{itemize}

The \(Store\) provides the low-level interface to the actual
memory-access operations.%
\footnote{%
This reflects the situation of
many modern systems, ranging from functional languages like ML or
Haskell to object-oriented languages like C\# or Java. In languages
like C or \Cpp the situation is more intricate.  } %
But on this abstract level its specification also provides the basic
terminology that is needed for talking about garbage collection. In
particular we use the following sets:

\begin{itemize}
\item \strut\(active\) are those nodes that constitute the Mutator's
graph.
\item \strut\(supply\) are the nodes in the freelist. (They become
\(active\) through the operation \(addNew\).)
\item \strut\(live\) is a shorthand for the union of the \(active\)
and \(supply\) nodes.
\item \strut\(dead\) are the garbage nodes that are neither reachable
from the Mutator nor in the freelist. (Nodes may become \(dead\)
through the operation \(delArc\).)
\end{itemize}

Note that the specifications in \Fig{fig:Architecture} use \(A = B
\uplus C\) as a shorthand notation for the two properties \(A = B \cup
C\) and \(B \cap C = \emptyset\). They also use overloading of
operation names. For example \(active\) is used both for the subgraph
that constitutes the Mutator's view and for the set of nodes in this
subgraph. Such overloaded symbols must always be distinguishable from
their context.

Note also that we frequently refer to the ``set'' \(Arcs\) of the arcs
of a graph and also to the ``set'' \(sucs(a)\) of all successors of a
node \(a\); but these are actually \emph{multisets}, since two nodes
may be connected by several arcs. (Technically, the cell has several
slots that all point to the same cell.)

\subsection{Fundamental Properties of the Mutator}
\label{sec:Mutator-basics}

The Mutator's operations \(addArc, delArc, addNew\) have an invariant
property that is decisive for the working of any kind of garbage
collector: \emph{being garbage is a stable property}
\cite{Azatchi+2003}.

\medskip

\begin{proposition}[Antitonicity of
Mutator]\label{prop:Antitone-Mutator} A Mutator can never ``escape''
the realm given by its graph and the freelist; that is, it can never
reclaim dead garbage nodes. In other words, the realm of live nodes
(graph + freelist) monotonically \emph{decreases}.
\end{proposition}

\subsection{The Fundamental Specification of Garbage Collection}
\label{sec:GC-Spec}

Surprisingly often papers on garbage collection refer to an intuitive
understanding of what the Collector shall achieve. But in a formal
treatment we cannot rely on intuition; rather we have to be absolutely
precise about the goal that we want to achieve.

Consider the architecture sketched in \Fig{fig:Architecture}. The
Mutator continuously performs its basic operations \(addArc, delArc\)
and \(addNew\), which -- from the Mutators viewpoint -- are all
considered to be total functions; i.e. they return a defined value on
all inputs. This is trivially so for \(addArc\) and \(delArc\), since
their arguments are existing in the mutators graph. The problematic
operation is \(addNew\), since this operation needs an element from
the freelist. However, the freelist may be empty (i.e. \(supply =
\0\)). In this situation there are two possibilities:

\begin{itemize}
\item \strut\(|active| = MemorySize\). That is, the Mutator has
used all available memory in its graph. Then nothing can be
done!%
\item \strut\(|active| < MemorySize\). When \(supply = \0\), this
means that \(dead \neq \0\).  This is the situation in which we want to
recycle dead garbage cells into the freelist. And this is the
Collector's reason for existence!
\end{itemize}

Based on this reasoning, we obtain two basic principles for the
Mutator/Collector paradigm.

\medskip
\begin{assumption}[Boundedness of Mutator's graph]\label{asmp:Bounded}
\strut\(|Mutator.graph| < MemorySize\)
\end{assumption}

Under this global assumption the Collector has to ensure that the
operation \(addNew\) is a total function (which may at most be
delayed). This can be cast into a temporal-logic formula:

\medskip
\begin{goal}[Specification of Collector]\label{goal:Collector}
\strut\([]\, <> supply \neq \emptyset\) \hspace{.5cm}(provided
assumption \ref{asmp:Bounded} holds)
\end{goal}

This is a liveness property stating that ``at any point in time the
freelist (may be empty but) will eventually be nonempty.''  When this
condition is violated, that is, \(supply = \0\), then it follows by
the global \Asmp{asmp:Bounded} that \(dead \neq \0\). Hence the
Collector has to find at least \emph{some} dead nodes, which it can
then transfer to the freelist.%
This can be cast into an operation \(recycle\) with the initial
specification given in \Fig{fig:Collector-Initial}.

\begin{figure}[htbp]
\begin{center}
\Unit{1mm}
\begin{spec}[box=\progA,width=72\unit]
recycle: Graph(Node,Arc) -> Set(Node)
\0 \subset recycle(G) \subseteq dead \quad\& if dead \neq \0
\0 = recycle(G) \quad\& if dead = \0
\end{spec}
\Spec Coll[0,0]={Collector}{\usebox\progA}
\caption{The Collector's task}
\label{fig:Collector-Initial}
\end{center}
\end{figure}


Hence we should design the system's working such that the following
property holds (using an ad-hoc notation for transitions).

\medskip
\begin{goal}[Required actions of
Collector]\label{goal:Collector-Action} \strut\([]\,<> \big(supply
\longrightarrow supply \uplus recycle(G)\big)\)
\end{goal}

When \Goal{goal:Collector-Action} is met, then the original
\Goal{goal:Collector} is also guaranteed to hold. In other words, the
collector has to periodically call \(recycle\) and add the found
subset of the garbage nodes to the freelist.

Note that the above operation can happen at any point in time; we need
not wait until the freelist is indeed empty. (This observation leaves
considerable freedom for optimized implementations which are all
correct.)

\subsection{How to Find Dead Nodes}
\label{sec:Find-dead-nodes}


Unfortunately, the specification of \(recycle\) in Figure
\ref{fig:Collector-Initial} is not easily implementable since the dead
nodes are not directly recognizable.  Since the $dead$ nodes are
computed by taking the complement of the $live$ nodes (i.e. $live =
\complement dead = nodes backslash live$), the idea comes to mind to
work with the complement of \(recycle\). This leads to the simple
calculation

\begin{spec}
\0 \subset recycle(G) \subseteq dead 
<=> \complement\, \0 \supset \complement\, recycle(G) \supseteq \complement\, dead
<=> nodes \supset \complement\, recycle(G) \supseteq live
<=> nodes \supset trace(G) \supseteq live
\end{spec}

where we introduce a new function \(trace(G)\) such that 
\(recycle(G) = \complement\, trace(G)\). 

This leads to the refined version of the Collector's specification in
\Fig{fig:Collector-V1}.

\begin{figure}[htbp]
\begin{center}
\Unit{1mm}
\begin{spec}[box=\progA,width=72\unit]
recycle: Graph(Node,Arc) -> Set(Node)
trace: Graph(Node,Arc) -> Set(Node)

recycle(G) = \complement\,trace(G)
live \subseteq trace(G) \subset nodes \quad\& if dead \neq \0
trace(G) = nodes \quad\& if dead = \0
\end{spec}
\Spec Coll[0,0]={Collector}{\usebox\progA}
\caption{The Collector's task (first refinement)}
\label{fig:Collector-V1}
\end{center}
\end{figure}

Note that this specification, which will form the starting point for
our more detailed derivation, is \emph{formally derived} from the
fundamental requirements for garbage collection as expressed in
\Asmp{asmp:Bounded} and \Goal{goal:Collector} above!

\subsection{An Intuitive Example and a Subtle Bug}
\label{sec:Example and Bug}

\begin{figure}[htbp]
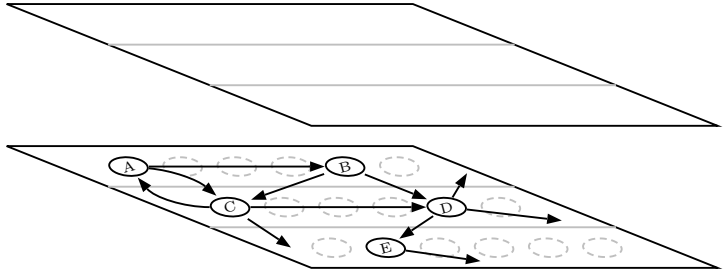

\vspace{-1\baselineskip}
\begin{center}
\Unit{1mm}
\psscalebox{.9}{%
\begin{gc-animation}
  \UpperPlane

  \Visible{A,B,C,D,E,F,G,H,I,J}
  \LowerPlane
\end{gc-animation}
}
\caption{At the start of the Collector}
\label{fig:Intuition-Start}
\end{center}
\end{figure}

We demonstrate the working of a typical garbage collection algorithm
by a simple example. \Fig{fig:Intuition-Start} illustrates the
situation at the beginning of the Collector by showing a little
fragment of the store; solid nodes are reachable from the root \(A\),
dashed circles represent dead garbage nodes (the arcs of which are not
drawn here for the sake of readability). We use the metaphor of
``planes'' to illustrate both mark-and-sweep and copying collectors.
In the former, ``lifting'' a node to the upper plane means marking, in
the latter it means copying. The picture already hints at a later
generalization, where the store is partitioned into ``regions''.

\begin{figure}[htbp]
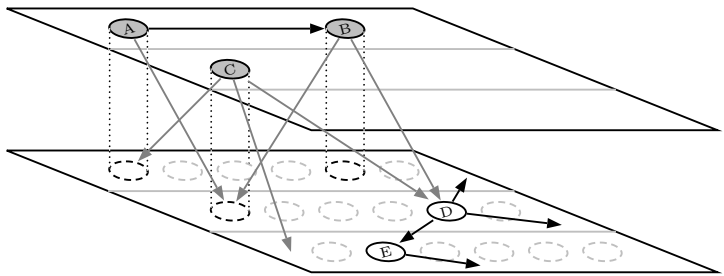

\vspace{-1\baselineskip}
\begin{center}
\Unit{1mm}
\psscalebox{.9}{%
\begin{gc-animation}
  \Visible{A,B,C}
  \Hot{A,B,C}
  \Down{AC,BC,BD,CA,CI,CD}
  \UpperPlane

  \Proxy{A,B,C}
  \Visible{D,E,F,G,H,I,J}
  \LowerPlane
\end{gc-animation}
}
\caption{A snapshot}
\label{fig:Intuition-Snapshot-1}
\end{center}
\end{figure}

\Fig{fig:Intuition-Snapshot-1} shows an intermediate snapshot of the
algorithm. Some nodes and arcs are already lifted (i.e. marked or
copied), others are still not considered. The gray nodes are in the
``hot zone'' -- the so-called ``workset'' --, which means that they
are marked/copied, but not all outgoing arcs have been handled yet.

\begin{figure}[htbp]
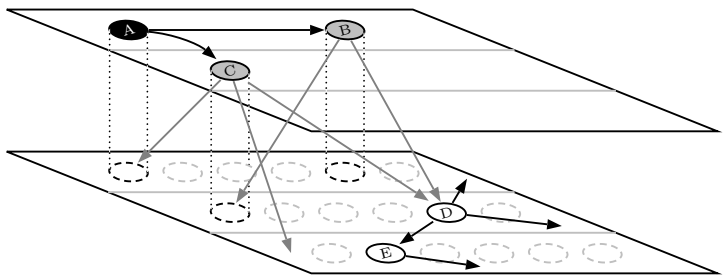

\vspace{-1\baselineskip}
\begin{center}
\Unit{1mm}
\psscalebox{.9}{%
\begin{gc-animation}
  \Visible{A,B,C}
  \Hot{B,C}
  \Black{A}
  \Down{BC,BD,CA,CI,CD}
  \UpperPlane

  \Proxy{A,B,C}
  \Visible{D,E,F,G,H,I,J}
  \LowerPlane
\end{gc-animation}
}
\caption{The next snapshot}
\label{fig:Intuition-Snapshot-2}
\end{center}
\end{figure}

\Fig{fig:Intuition-Snapshot-2} shows the next snapshot. Now all direct
successors of \(A\) have been treated. Therefore \(A\) is taken out of
the workset -- which we represent by the color black.

Note that we have the invariant property (which will play a major role
in the sequel) that all downward arrows start in the workset. This
corresponds to one of the two main invariants in the original paper of
Dijkstra et al.~\cite{Dijkstra+1978}.

\begin{figure}[htbp]
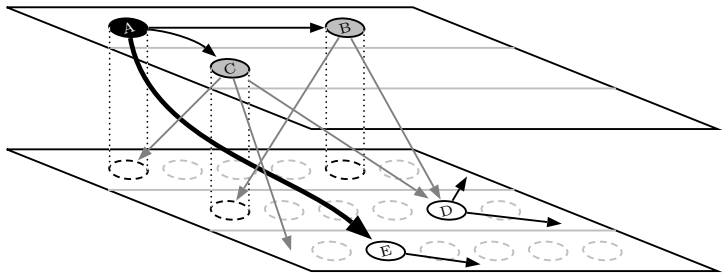

\vspace{-1\baselineskip}
\begin{center}
\Unit{1mm}
\psscalebox{.9}{%
\begin{gc-animation}
  \Visible{A,B,C}
  \Hot{B,C}
  \Black{A}
  \Down{BC,BD,CA,CI,CD}
  \Invisible{DE}
  \UpperPlane

  \Proxy{A,B,C}
  \Visible{D,E,F,G,H,I,J}
  \LowerPlane
  \nccurve[angleA=280,angleB=140,linecolor=black,linewidth=2pt,fillstyle=none]{->}{A}{E}
\end{gc-animation}
}
\caption{A subtle error}
\label{fig:Intuition-Problem}
\end{center}
\end{figure}

\emph{A subtle problem}: Now let us assume that in this moment the
Mutator intervenes by adding an arc \(A -> E\) and then deleting the
arc \(D -> E\). This leads to the situation depicted in
\Fig{fig:Intuition-Problem}.  Since \(A\) is no longer in the workset,
its connection to \(E\) will not be detected. Hence, \(E\) is
\emph{hidden from the Collector} \cite{Vechev+2006} and therefore will
be treated as a dead garbage node -- which is a severe bug!

\emph{Any formal method for deriving garbage collection algorithms
must ensure that this bug cannot happen!}
Note that this situation violates the invariant about the downward
arrows. And our formal treatment will show that keeping this invariant
intact is actually the clue to the derivation of correct garbage
collectors.

There are three reasonable ways to cope with this problem (using
suitable \emph{write barriers}):
\begin{itemize}
\item When performing \(addArc(A,E)\), record \(E\). (This is the
approach of Dijkstra et al.~\cite{Dijkstra+1978}.)
\item When performing \(addArc(A,E)\), record \(A\). (This is the
approach taken by Steele \cite{Steele:1975}.)
\item When performing \(delArc(D,E)\), record \(E\). (This is the
approach taken by Yuasa \cite{Yuasa:1990}.)
\end{itemize}

Vechev et al.~\cite{Vechev+2006} speak of ``installation-protected''
in the first two cases and of ``deletion-protected'' in the last case.

Note also that this bug may appear in an even subtler way
\emph{during} the handling of a node in the workset. Consider the node
\(C\) in \Fig{fig:Intuition-Snapshot-2} and suppose that it has lifted
the first two of its three arcs. At this moment the Mutator redirects
the first pointer field to, say, \(E\). But a naive Collector will
nevertheless take node \(C\) out of the workset (color it black) when
its final arc has been treated. -- The same bug again!

\medskip

At this point we will leave the concrete considerations about garbage
collectors and pass on to the more abstract viewpoint of fixed points
and lattices (or cpos). In the terminology of \Fig{fig:Pushout} we
follow the upward arrow, that is, we generalize a concrete problem
into a more abstract one. Once this is done, we can derive a whole
variety of solutions in a strictly top-down fashion.





\section{Mathematical Foundation: Fixed Points}
\label{sec:Fixpoints}

In garbage collection one can roughly distinguish two classes of collectors (see \Sec{sec:Sequential-concurrent}):
\begin{itemize}
\item \emph{Stop-the-world collectors}: these are the classical
  non-concurrent collectors, where the mutators need to be stopped,
  while the collector works.
\item \emph{Concurrent collectors}: these are the collectors that
  allow the mutators to keep working concurrently with the collector
  (except for very short pauses).
\end{itemize}

As we will demonstrate in a moment, the traditional stop-the-world
collectors correspond on the abstract level to classical fixed-point
theory. For the concurrent collectors we need to generalize this
classical fixed-point theory to a variant that we baptized ``dynamic
fixed points''.

We briefly review the classical theory before we present our
generalization.

\subsection{Classical Fixed Points (Stop-the-world Collectors)}
\label{sec:Classical-Fixpoint}

The best known treatments of the classical fixpoint problem in
complete lattices are those of Tarski \cite{Tarski:1955} and Kleene
{\cite{Kleene56}}.  Before we quote these we present some relevant
terminology (assuming that the reader is already familiar with the
very basic notions of partial order, join, meet etc.)

\begin{itemize}
\item For a set \(s = { x_0, x_1, x_2, ...}\) of type \(Set(A)\) and a
  function \(f: A -> A\) we use the overloaded function \(f: Set(A) ->
  Set(A)\) by writing \(f(s)\) as a shorthand for \({f(x_0), f(x_1),
    f(x_2), ...}\). (In functional-programing notation this would be
  written with the apply-to-all operator as \(f*s\).)

\item A function \(f: A -> A\) is \emph{monotone}, if \(x =< y => f(x)
=< f(y)\) holds.

\item The function \(f\) is \emph{continuous}, if \(f(|_| {x_0, x_1,
x_2, ... }) = |_| { f(x_0), f(x_1), f(x_2), ... }\) holds. This could
be shortly written as \(f \o |_| = |_| \o f\) (by using the overloaded
versions of the symbol \(f\)).

\item The function \(f\) is \emph{inflationary} in \(x\), if \(x =<
f(x)\) holds. Then \(x\) is called a \emph{post-fixed point} of
\(f\). (Analogously for \emph{pre-fixed points}.)

\item The element \(x\) is called a \emph{fixed point} of \(f\), if
\(x = f(x)\) holds; \(x\) is the \emph{least fixed point}, if \(x =<
y\) for any other fixed point \(y\) of \(f\).

\item The element \(x\) is called a \emph{fixed point} of \(f\)
\emph{relative} to \(r\), if \(x = f(x) /\ r =< x\) holds.

\item By \(\fh(x) = LEAST u.\ u = f(u) /\ x =< u\) we denote the
  reflexive-transitive \emph{closure} of \(f\) (when it exists); i.e.
  the function that yields the least fixed point of \(f\) relative to
  \(x\).
\end{itemize}

\begin{lemma}[Properties of the closure \(\fh\)]\label{lemma:Closure-properties}
The closure \(\fh(x)\) has a number of properties that we will utilize frequently:
\begin{itemize}
\item \strut\(x =< \fh(x)\) (inflationary);
\item \strut\(\fh(\fh(x)) = \fh(x)\) (idempotent);
\item \strut\(f(\fh(x))=\fh(x)\) (fixpoint);
\item \strut\(\fh(f(x)) = \fh(x)\) if \(x =< f(x)\)
\end{itemize}

\end{lemma}

\emph{Proof.} The first three properties follow directly from the
definition of \(\fh\). The last one can be seen as follows: Denote \(u
= \fh(x)\) and \(v = \fh(f(x))\). Then we have by monotonicity

\quad \(x =< f(x) |- \fh(x) =< \fh(f(x)) |- u =< v\).

On the other hand we have

\quad \(x =< u |- f(x) =< f(u) = u\).

Since \(v\) is the least value with \(v = f(v) \wedge f(x) =< v\), we have
\(v =< u\). \Eop

\medskip

\begin{theorem}[Tarski]\label{theo:Tarski}
  Let $L$ be a complete lattice and $f: L \rightarrow L$ a monotone
  function on $L$. Then $f$ has a complete lattice of fixed points. In
  particular the least fixed point is the meet of all its pre-fixed
  points and the greatest fixed point is the join of all its
  post-fixed points.
\end{theorem}

\medskip

\begin{theorem}[Kleene]\label{theo:Kleene}
  For a \emph{continuous} function \(f\) the least fixed point \(x\)
  is obtained as the least upper bound of the Kleene chain:
	\begin{center}
	  \strut\(x = |_| { _|_, f(_|_), f^{2}(_|_), f^{3}(_|_), ... }\)
	\end{center}
	where \(_|_\) is the bottom element of the lattice.
\end{theorem}

\medskip

In the meanwhile it has been shown that the essence of these theorems
also holds in the simpler structure of \emph{complete partial orders
  (cpos)}\footnote{A {\em cpo} is a partial order in which every
  directed subset has a supremum}.

\medskip


More recently Cai and Paige \cite{CaiPaige:1989} published a number of
generalizations that are streamlined towards practical algorithmic
implementations of fixpoint computations. We paraphrase their main
result here, since we are going to utilize it as a ``blueprint'' for
our subsequent development.

\medskip

\begin{theorem}[Cai-Paige]\label{theo:Paige}
  Let A be a cpo and \(f: A -> A\) be a monotone function that is
  inflationary in \(r\). Let moreover \strut\({s_{0}, s_{1}, s_{2},
    ..., s_{n} }\) be an arbitrary sequence obeying the conditions
\begin{center}
\strut\(r = s_{0} \) \\
\strut\(s_{i} < s_{i+1} =< f(s_{i})\) \hbox to0pt{\quad for \(i < n\)\hss}

\strut\(s_{n} = f(s_{n})\)
\end{center}
Then \(s_{n}\) is the \emph{least fixed point} of \(f\) relative to
\(r\). Conversely, when the least fixed point is finitely computable,
then the sequence will lead to such an \(s_{n}\).
\end{theorem}

Theorem \ref{theo:Paige} provides a natural abstraction from
workset-based iterative algorithms, which maintain a workset of change
items.  At each iteration, a change item is selected and used to
generate the next element of the iteration sequence.  The incremental
changes tend to be small and localized, hence this is called the
\emph{micro-step} approach and the Kleene chain the \emph{macro-step}
approach \cite{PH:2006}.  All practical collectors use a workset that
records nodes that await marking.

To illustrate these basic results, we derive the overall structure of
a stop-the-world collector.  The essence of it is the iterative
algorithm for finding garbage nodes to recycle.


Letting $roots$ denote the roots of the active graph together with the
head of the {\em supply} list, we have
$$live = \fh(roots)$$
where
$$f(R) = \{b\;|\; b\in G.sucs(a) \And a\in R\};$$
in words, the {\em active} nodes are the closure of the roots under the
successor function in the current graph $G$.

To derive an algorithm for computing the {\em dead} nodes, we
calculate as follows:

\begin{tabbing}
\= xxx\= xxx\= xxx\= xxx\= xxx\= xxx\= xxx\= \kill
\InitialStep{$dead$}

\CompactStep{=}{definition}{$\complement\; live$ }

\CompactStep{=}{definition}{$\complement\; \fh(roots)$}

\CompactStep{=}{using the law $\complement\; \hh(R)\Equals \ic(R)$ 
                      where $i(x)=\complement  h(\complement\;  x)$} 
   {$\gc (roots)$}
 
\end{tabbing}

where $\gc(R)$ is the greatest fixpoint of the monotone function
$$g(x) = nodes\setminus (roots \cup \{b\;|\; b\in sucs(a) \And a\in nodes\setminus x \}).$$

This allows us to produce a correct, but naive iterative algorithm to
compute dead nodes which is based on Theorem \ref{theo:Kleene}.

\begin{program}
\caption{Raw Fixpoint Iteration Program}
\label{alg:FP0}

\begin{codebox}
\li $W \gets h.nodes;$  
\li \While $W \neq g(W)$ do $W \gets g(W)$
\li \Return $W$
\end{codebox}

\end{program}

Following Cai and Paige \cite{CaiPaige:1989}, we can construct an efficient
fixpoint iteration algorithm using a workset defined by
$$WS \Equals \WSvar \setminus g(\WSvar).$$ 
Although this workset definition is created by instantiating a
problem-independent scheme, it has an intuitive meaning: the workset
is the set of nodes whose parents have been ``marked'' as live, but
who themselves have not yet been marked.  The workset expression can
be simplified as follows
\begin{tabbing}
xxxxx\= xxx\= xxx\= xxx\= xxx\= xxx\= xxx\= xxx\= \kill
\InitialStep{$\WSvar \setminus g(\WSvar)$}

\MiddleStep{=}{Definition}
	{$\WSvar \setminus ( nodes \setminus (roots \cup \{b\;|\; b\in sucs(a) 
                                               \And a\in nodes \setminus \WSvar \}))$}

\MiddleStep{=}{Using the law $A\setminus (B\cup C) \Equals (A\setminus B)\setminus C$ }
     {$\WSvar \setminus (( nodes \setminus roots) 
         \setminus \{b\;|\; b\in sucs(a)\And a\in nodes \setminus \WSvar \})$}

\MiddleStep{=}{Using the law $A\setminus (B \setminus C)
               \Equals (A\setminus B) \Union (A\cap C)$ }
     {$(\WSvar \setminus ( nodes \setminus roots)) 
         \Union (\{b\;|\; b\in sucs(a)\And a\in nodes \setminus \WSvar \}\cap \WSvar)$}

\MiddleStep{=}{Using the law $\{x|P(x)\}\cap Q \Equals \{x|P(x)\wedge x\in Q\}$ }
     {$(\WSvar \setminus ( nodes \setminus roots)) 
         \Union \{b\;|\; b\in sucs(a)\And b\in \WSvar \And a\in nodes \setminus \WSvar \}$}

\MiddleStep{=} {Again using the law $A\setminus (B \setminus C)
               \Equals (A\setminus B) \Union (A\cap C)$ (on first term)}
     {$(\WSvar \setminus nodes) \cup (\WSvar \cap roots) 
         \Union \{b\;|\; b\in sucs(a)\And b\in \WSvar \And a\in nodes \setminus \WSvar \}$}

\MiddleStep{=} {Simplifying}
     {$\{\} \cup (\WSvar \cap roots) 
         \Union \{b\;|\; b\in sucs(a)\And b\in \WSvar \And a\in nodes \setminus \WSvar \}$}

\FinalStep{=} {Simplifying}
          {$(\WSvar \cap roots) \Union 
            \{b\;|\; b\in sucs(a)\And b\in \WSvar \And a\in nodes \setminus \WSvar \}
	   $}
\end{tabbing}

The greatest fixpoint expression can be computed by the workset-based
Program \ref{alg:FP1}, which is based on Theorem \ref{theo:Paige}.

\begin{program}
\caption{Workset-based Fixpoint Iteration Program}
\label{alg:FP1}

\begin{codebox}
\li $W \gets nodes;$  
\li \While $\exists z \in ((W \cap roots) \Union 
            \{b\;|\; b\in sucs(a)\And b\in W 
	     \And a\in nodes \setminus W \}\})$ 
\li \qquad $W \gets W - z$
\li \Return $W$
\end{codebox}

\end{program}

To improve the performance of this algorithm, we apply the Finite
Differencing transformation \cite{Paige8207}, according to which we
incrementally maintain the invariant 
$$ WS = (W \cap roots) \Union
  \{b\;|\; b\in sucs(a)\And b\in W \And a\in nodes \setminus W \}.$$

There are two places in the code that might disrupt the invariant, in
lines 1 and 3.  We maintain the invariant in line 1 with respect the
initialization $W=nodes$ as follows:
\begin{tabbing}
Assumexx\= xxx\= xxx\= xxx\= xxx\= xxx\= xxx\= xxx\= \kill
\Assume{$W=nodes$} \\
\Simplify{$WS = (W \cap roots) \Union
          \{b\;|\; b\in sucs(a)\And b\in W \And a\in nodes \setminus W \} $} \\

 \\
WS=\= xxx\= xxx\= xxx\= xxx\= xxx\= xxx\= xxx\= \kill

\InitialStep{$(WS = (roots \cap nodes) \Union 
   \{b\;|\; b\in sucs(a)\And b\in nodes \And a\in nodes \setminus nodes \}$}

\CompactStep{=}{ }
     {$roots \Union 
   \{b\;|\; b\in sucs(a)\And b\in nodes \And a\in \{\} \}$}

\CompactStep{=}{}{$roots \Union \{\}$}

\CompactFinalStep{=}{}{$roots$}
\end{tabbing}

and incremental code (wrt the change $W' \Equals W - z$):
\begin{tabbing}
Assumexx\= xxx\= xxx\= xxx\= xxx\= xxx\= xxx\= xxx\= \kill
\Assume{$WS = (W \cap roots) \Union
          \{b\;|\; b\in sucs(a)\And b\in W \And a\in nodes \setminus W \}$\\
	\> $\And W' \Equals W - z$} \\
\Simplify{$WS' = (W' \cap roots) \Union
          \{b\;|\; b\in sucs(a)\And b\in W' \And a\in nodes \setminus W' \}$} \\

xxxxx\= xxx\= xxx\= xxx\= xxx\= xxx\= xxx\= xxx\= \kill
 \\
\InitialStep{$W' \cap roots) \Union 
   \{b\;|\; b\in sucs(a)\And b\in W' \And a\in nodes \setminus W' \}$}

\MiddleStep{=}{Using assumption $W' \Equals W - z$ }
     {$((W - z)  \cap roots) \Union 
              \{b\;|\; b\in sucs(a)\And b\in (W - z) \And a\in nodes 
              \setminus (W - z) \}$}

\MiddleStep{=}{Simplifying} 
     {$((W \cap roots) - z) \Union 
              (\{b\;|\; b\in sucs(a)\And b\in W \And a\in nodes 
              \setminus (W - z) \}  - z)$}

\MiddleStep{=}{Pulling out common subtraction of $z$} 
     {$(W \cap roots) \Union 
              \{b\;|\; b\in sucs(a)\And b\in W \And a\in nodes 
              \setminus (W - z) \}  - z$}

\MiddleStep{=}{distribute element membership} 
     {$(W \cap roots) \Union 
              \{b\;|\; b\in sucs(a)\And b\in W 
                   \And (a\in (nodes \setminus W) \Or a=z) \}  - z$}

\MiddleStep{=}{distribute set-former over disjunction} 
     {$(W \cap roots)$\\ 
       \>\> $\Union 
              (\{b\;|\; b\in sucs(a)\And b\in W 
                   \And a\in (nodes \setminus W) \}$\\ 
       \>\> $\Union \{b\;|\; b\in sucs(a)\And b\in W \And  a=z) \})
	       - z$}

\FinalStep{=}{fold definition of $WS$, and simplify} 
     {$WS \Union \{b\;|\; b\in sucs(z)\And b\in W \}) - z$}

\end{tabbing}

The resulting code is shown in Program \ref{alg:FP2}.

\begin{program}
\caption{Optimized Fixpoint Iteration Program}
\label{alg:FP2}

\begin{tabbing}
xxxxxxx\= xxx\= xxx\= xxx\= xxx\= xxx\= xxx\= xxx\= \kill
$\text{invariant} \; 
   WS = (W \cap roots) \Union 
        \{b\;|\; b\in h.sucs(a)\And b\in W \And a\in h.nodes \setminus W \}$ \` 1 \\
$W\; := \; h.nodes \;||\; WS\; := \; roots; $ \` 2 \\
$\text{while}\; \exists z \in WS\; \text{do}$\` 3 \\
\> $W\; := W - z\ 
  \;||\; WS \; :=\; WS \Union \{b\;|\; b\in h.sucs(z)\And b\in W \}) - z$ \` 4 \\
$\text{output} \; W.$ \` 5
\end{tabbing}
\end{program}



Programs \ref{alg:FP1} and \ref{alg:FP2} represent the abstract
structure of most marking algorithms.  Our point is that its
derivation, and further steps toward implementation, are carried out
by generic, problem-independent transformations, supported by
domain-specific simplifications, as above.

Further progress toward a detailed implementation requires a variety
of other transformations, such as finite differencing, simplification,
and datatype refinements.  For example, the finite set $W$ may be
implemented by a characteristic function, which in turn is refined to
a bit array, or concurrent data structures for local buffers or
work-stealing queues.


As our final preparatory step within the realm of the classical
fixed-point concepts we mention a central property that is the core of
the correctness proof for implementations. If we compute the sequence
\(s_{0}, s_{1}, s_{2}, ..., \) by a loop, then a Hoare-style
verification would need the following invariant.

\medskip

\begin{corollary}[Invariance of closure]\label{prop:Closure-invariance}
  The elements of the set \({ s_{0} < s_{1} < s_{2} < ... < s_{n} }\)
  all have the same closure:
\begin{center}
\strut\(\fh(s_{i}) = \fh(r)\)
\end{center}
\end{corollary}

\emph{Proof.} This invariance follows directly from monotonicity and
the properties of \(\fh\) stated in \Lemma{lemma:Closure-properties}:
\(\fh(s_{i}) =< \fh(s_{i+1}) =< \fh(f(s_{i})) = \fh(s_{i})\). \Eop

\subsection{Fixed Points in Dynamic Settings
 (Concurrent Collectors)}
\label{sec:Dynamic-Fixpoint}

The classical fixed-point considerations work with a fixed monotone
function \(f\).  In the garbage collection application this is
justified as long as the graph, on which the collector works, remains
fixed during the collector's activities. But as soon as the mutator
keeps working in parallel with the collector, the graph keeps
changing, while the collector is active. This can be modeled by
considering a sequence of graphs \(G_{0}, G_{1}, G_{2}, ...\) and by
making the function \(f\) dependent on these graphs: \(f(G_{0})(...),
f(G_{1})(...), f(G_{2})(...), ...\), where \(f: Graph\rightarrow
Set(Node)\rightarrow Set(Node)\) and
$$f(G)(S)= S\Union \{b\;|\; a\in S \And b \in G.sucs(a)\}.$$ 
Intuitively, $f$ extends a given set of nodes with the set of their
successors in the graph. To ease readability we omit the explicit
reference to the graphs and simply write \(f_{0}, f_{1}, f_{2}, ...\).

\subsubsection{The foundation}
\label{sec:Foundation}

Using this notational liberty the specification of the underlying
foundation is stated in \Fig{fig:Foundation}: the \(f_{i}\) are
monotone \Mon and inflationary in \(r\) \Inf. Moreover the
closure-forming operator \(\fh\) is defined by \Fp.

\begin{spec}[plain,box=\progA,width=.85\textwidth]
EXTEND Cpo(A)		\- $A$ is a cpo (alternatively: lattice)
f_{0}, f_{1}, f_{2}, ... : A -> A	      \- sequence of functions
\widehat{_}: A -> A -> A -> A			  \- $\fh$ is reflexive-transitive closure of $f$
r: A								           \- ``root''

x =< y => f_{i}(x) =< f_{i}(y) \& \Mon		\- all $f\sb{i}$ are monotone\hfill
r =< f_{i}(r)	                 \& \Inf		\- all $f\sb{i}$ are inflationary in $r$

\fh(x) = LEAST s:\ x =< s /\ s = f(s) \qquad\& \Fp      \- closure (computes least fixed point)
\end{spec}%
\begin{figure}[htbp]%
\begin{center}%
\Unit{0.01\linewidth}%
\Spec InitialSpec[0,29]={Foundation}{\usebox\progA}%
\caption{Initial Specification}%
\label{fig:Foundation}%
\end{center}%
\end{figure}%
%

\subsubsection{The initial problem formulation}
\label{sec:Initial-problem}

Based on this foundation we can now formulate our goal. Recall the
specification of the garbage collection task given by \emph{Collector}
in \Fig{fig:Collector-V1}: \(live \subseteq trace(G) \subset nodes\).
This translates into our dynamic setting as 
\(live_{n} \subseteq s  \subset nodes\).
We add as a working hypothesis that the set
\(live_{0}\) serves as an upper bound that we will need to guarantee in our 
dynamic algorithm:
\(live_{n} \subseteq s \subseteq live_{0} \subset nodes\).

The set \(live_{0}\) is sometimes called the
``snapshot-at-the-beginning'' \cite{Azatchi+2003}.  Since in our
abstract setting \(live_{n}\) corresponds to the closure
\(\fh_{n}(r)\) and \(live_{0}\) corresponds to the closure
\(\fh_{0}(r)\), we immediately obtain the abstract formulation \Aim of
our problem statement (\Fig{fig:InitialSpec}).

\begin{spec}[box=\progA,width=.85\textwidth]
EXTEND Foundation 
r =< x => f_{i+1}(\fh_{i}(x)) =< \fh_{i}(x)  \qquad\qquad \& \Caged \- garbage can only grow

THM\ \exists\, n,\, s:\ \fh_{n}(r) =< s =< \fh_{0}(r) \qquad\qquad\&\Aim \- $\mathit{live}\sb{n} \leq s \leq \mathit{live}\sb{0}$

THM r =< x => \fh_{0}(x) >= \fh_{1}(x) >= \fh_{2}(x) >= ...\qquad\qquad\& \Anti \- Lemma \ref{lemma:antitone-closure}
\end{spec}%

\begin{figure}[htbp]
\begin{center}%
\Unit{0.01\linewidth}
\Spec InitialSpec[0,19]={Fixpoint-Problem}{\usebox\progA}  
\caption{Initial Specification}
\label{fig:InitialSpec}
\end{center}
\end{figure}

Axiom \Caged is the abstract counterpart of the fundamental
\Prop{prop:Antitone-Mutator} in \Sec{sec:Mutator-basics}: the set of
live nodes is monotonically decreasing over time, or, dually, garbage
increases monotonically.  For proof-technical reasons we have to
conditionalize this property to any set \(x\) containing the roots
\(r\).)%

Note that the existential formula \Aim is trivially provable by
setting \(n = 0\) and \(s=\fh_{0}(r)\). Actually the property \Anti
(see \Lemma{lemma:antitone-closure} below) shows that such an \(s\)
exists for any \(n\). However, our actual task will be to come up with
a \emph{constructive algorithm} that yields such an \(n\) and \(s\).%
 \footnote{%
   It may be interesting to compare property \Aim to the classical
   non-concurrent situation expressed in
   Lemma~\ref{prop:Closure-invariance}. Instantiated to the final value
   $s = s_{n}$ Lemma~\ref{prop:Closure-invariance} states $s = \fh(r).$
   This equality can be formally rewritten into the two inequalities
   \(\fh(r) =< s =< \fh(r)\).  This is weakened in property \Aim by
   setting on the left \(\fh \ \widehat{=}\ \fh_{n}\) and on the right
   \(\fh\ \widehat{=}\ \fh_{0}\).  Similar kinds of weakenings are
   also found in Hoare- or Dijkstra-style program developments, when
   deriving invariants from given pre- or post-conditions.
 }

For the specification \(FixpointProblem\) we can prove the property
\Anti (i.e.~\Lemma{lemma:antitone-closure}) that will be needed later
on. This monotonic decreasing of the closure is in accordance with our
intuitive perception of the Mutator's activities. The operation
\(delArc\) may lead to fewer live. And the operations \(addArc\) and
\(addNew\) do not change the set of live nodes (since the freelist is
part of the live nodes).

\medskip
\begin{lemma}[Antitonicity of closure]\label{lemma:antitone-closure}
The closures are monotonically decreasing:
\begin{center}
For\quad \(r =< x\)\quad we have\quad \(\fh_{0}(x) >= \fh_{1}(x) >= \fh_{2}(x) >= ...\) \qquad\Anti
\end{center}
\end{lemma}

\emph{Proof}: We use a more general formulation of this lemma: For
monotone \(g\) and \(h\) we have the property
\begin{center}
\strut\(\forall x:\ g(\hh(x)) =< \hh(x) => \gh(x) =< \hh(x)\)
\end{center}
We show by induction that \(\forall i:\ g^{i}(x) =< \hh(x)\).
Initially we have \(g^{0}(x) = x =< \hh(x)\) due to the general
reflexivity property \Fp of the closure. The induction step uses the
induction hypothesis and then the premise: \(g^{i+1}(x) = g(g^{i}(x))
=< g(\hh(x)) =< \hh(x)\).

By instantiating \(f_{i+1}\) for \(g\) and \(f_{i}\) for \(h\) we
immediately obtain \(\fh_{i+1}(x) =< \fh_{i}(x)\) by using the axiom
\Caged, when \(r =< x\).
\Eop

\subsection{The \emph{Microstep} Refinement}
\label{sec:Microstep-refinement}

In order to get closer to constructive solutions we perform our first
essential refinement. Generalizing the idea of Cai and Paige (see
\Theo{theo:Paige}) we add further properties to our specification,
resulting in the new specification of \Fig{fig:Microstep-Spec}. Note
that we now use some member \(s_{n}\) of the sequence \(s_{0}, s_{1},
s_{2}, ...\) as a witness for the existentially quantified \(s\).

\begin{spec}[box=\progA,width=.85\textwidth]
EXTEND FixpointProblem
s_{0}, s_{1}, s_{2}, ... : A	      \- sequence of approximations

s_{0} = r					      \& \Start              \- start with ``root''
s_{i} < s_{i+1} =< f_{i}(s_{i}) \/ s_{i} = f_{i}(s_{i})\qquad\& \Step \- computation step

THM\ \exists\, n:\ \fh_{n}(r) =< s_{n} =< \fh_{0}(r) \qquad\& \AimN \- to be shown below
THM\ \fh_{0}(s_{0}) >= \fh_{1}(s_{1}) >= ... >= \fh_{n}(s_{n}) \qquad\& \Decr \- Lemma \ref{lemma:Decrease-Closures} below
\end{spec}%
\begin{figure}[htbp]
\begin{center}%
\Unit{0.01\linewidth}
\Spec InitialSpec[0,25]={Micro-Step}{\usebox\progA}
\caption{The ``micro-step approach''}
\label{fig:Microstep-Spec}
\end{center}
\end{figure}


\emph{Proof of property \AimN}: In a finite lattice the \(s_{i}\)
cannot grow forever. Therefore there must be a fixpoint \(s_{n} =
f_{n}(s_{n})\) due to axiom \Step.
Then the left half of the proof of \AimN follows trivially from monotonicity:
\begin{spec}[width=.75\textwidth]
\forall i:\ r =< s_{i} 		\- axiom \Start and \Step
|- \ r =< s_{n} = f_{n}(s_{n})       \- $s\sb{n}$ is fixpoint
|- \ \fh_{n}(r) =< \fh_{n}(f_{n}(s_{n})) = \fh_{n}(s_{n}) = s_{n} \- properties of $\fh\sb{n}$ \Lemma{lemma:Closure-properties}
\end{spec}%
The right half \(s_{n} =< \fh_{0}(r)\) is a direct consequence of the following \Lemma{lemma:Decrease-Closures}. \Eop

\bigskip

\begin{lemma}[Decreasing Closures]\label{lemma:Decrease-Closures}
As a variation of \Lemma{lemma:antitone-closure} we can show property \Decr: the closures are \textbf{de}creasing, even when applied to the \textbf{in}creasing \(s_{i}\):
\begin{center}
\strut\(\forall i:\ \fh_{i+1}(s_{i+1}) =< \fh_{i}(s_{i})\)
\end{center}
\end{lemma}

\emph{Proof}: On the basis of \Lemma{lemma:antitone-closure} (property
\Anti in \Fig{fig:InitialSpec}) the proof follows directly from axiom
\Step by monotonicity:
\begin{spec}[width=.95\textwidth]
s_{i+1} =< f_{i}(s_{i})   						\- axiom \Step
|- \ \fh_{i+1}(s_{i+1}) =< \fh_{i+1}(f_{i}(s_{i})) =< \fh_{i}(f_{i}(s_{i})) = \fh_{i}(s_{i}) \- monotonicity of $\fh\sb{i+1}$; \Anti
\end{spec}
Note that \Anti is applicable here, since -- due to \Step{} -- \(r =< f_{i}(s_{i})\) holds.
\Eop

Lemma \ref{lemma:Decrease-Closures} may be depicted as follows:

\begin{center}
\Unit{1.3mm}
\begin{pspicture}(0,0)(58,34)
	
	\psellipse*[linecolor=red](10,17)(10,7)
	\psellipse*[linecolor=blue](06,17)(6,3)
	\psellipse*[linecolor=black](02,17)(2,1)
	
	\psellipse[linecolor=red](18,17)(18,12)
	\psellipse[linecolor=blue](24,17)(24,15)
	\psellipse[linecolor=black](29,17)(29,17)
	
	\white
	\rput(02,17){$s_{0}$}
	\rput(06.5,18.2){$s_{1}$}
	\rput(13.5,20){$s_{2}$}
	
	\rput(26,22){\red$\fh_{2}(s_{2})$}
	\rput(41,21){\blue$\fh_{1}(s_{1})$}
	\rput(53,20){\black$\fh_{0}(s_{0})$}
	
	\psline[linecolor=white]{->}(4,17)(20,17)
	\psline[linecolor=black]{<-}(36,17)(58,17)
	
\end{pspicture}
\end{center}

As can be seen here, the approximations \(s_{0}, s_{1}, s_{2}, ...\) keep growing, while at the same time their closures \(\fh_{0}(s_{0}), \fh_{1}(s_{1}), \fh_{2}(s_{2}), ...\) keep shrinking.

\emph{Remark}: This situation can also be rephrased as follows: We have a function \(F(f_{i}, s_{i})\) that is applied to the elements of two sequences. This function is antitone in the first argument and monotone in the second argument; we have to show that -- under the constraints given in our specification -- the function still is monotonically increasing.

\medskip

This essentially concludes the derivation that can reasonably be done on this highly abstract mathematical level of fixed points and lattices. However, in the literature one can find a variant of collectors, the development of which is best prepared on this level of abstraction as well.

\subsection{A Side Track: \emph{Snapshot Algorithms}}
\label{sec:Snapshots}

Consider again the specification \(Microstep\) in
\Fig{fig:Microstep-Spec}, where the goal is described in axiom \AimN
as \(\exists\, n:\ \fh_{n}(r) =< s_{n} =< \fh_{0}(r)\). Evidently
computing the value \(s_{n} = \fh_{0}(r)\) is an admissible solution.%

\footnote{%
  Remember that the closure computes the live nodes; axiom \AimN
  therefore means \(live_{n}\subseteq s_{n}\subseteq live_{0}\), which
  can be solved by \(s_{n} = live_{0}\). In other words, we compute
  the nodes that were live, when the Collector started.
} 
This approach, which has been used by Yuasa \cite{Yuasa:1990} and was
refined later by Azatchi et al.~\cite{Azatchi+2003}, is also referred
to as \emph{snapshot-at-the-beginning}.%

In order to follow this development path we refine the specification
\(Microstep\) in \Fig{fig:Microstep-Spec} to the specification
\(Snapshot\) in \Fig{fig:Snapshot} by requiring the additional
constraint \Snapshot that needs to be respected by later
implementations.
Note that the new axiom \Snapshot is the classical invariant that is
also used in non-concurrent garbage collectors.

\begin{spec}[box=\progA,width=.85\textwidth]
EXTEND MicroStep

\forall\, i:\ \fh_{0}(s_{i}) = \fh_{0}(r) \qquad\qquad\&\Snapshot  \- classical invariant $\mathit{live}\sb{i} = \mathit{live}\sb{0}$
\end{spec}%
\begin{figure}[htbp]
\begin{center}%
\Unit{0.01\linewidth}
\Spec InitialSpec[0,12]={Snapshot}{\usebox\progA}
\caption{The ``snapshot approach''}
\label{fig:Snapshot}
\end{center}
\end{figure}

\emph{How can this kind of computation be achieved in practice?} This
is demonstrated by Azatchi et al.~\cite{Azatchi+2003} based on a
technique that has been developed by some of the authors for a
concurrent reference-counting collector \cite{LevPet:2001}. Starting
from the fictitious idea of making a virtual snapshot by cloning the
complete original heap, it is then shown that this copying can be done
lazily such that only those nodes are actually copied that are
critical.

We only sketch this idea here abstractly in our framework: We
introduce a structure \(clone_{i}\) and an operator \(\nabla\) such
that \((G_{i}\nabla clone_{i}) \simeq G_{0}\); i.e. \(x\notin
clone_{i} => G_{i}.sucs(x) = G_{0}.sucs(x)\) and \(x\in clone_{i} =>
clone_{i}.sucs(x) = G_{0}.sucs(x)\).
The computation of the sequence \(s_{0}, s_{1}, s_{2}, ...\) is then
done based on \((G_{i}\nabla clone_{i})\) such that we effectively
always apply \(f_{0}\).

It remains to determine the structure \(clone_{i}\). There are
solutions of varying granularity, but the most reasonable choice
appears to be the following: Whenever the Mutator executes one of the
operations \(addArc(a,b)\), \(delArc(a,b)\) or \(addNew(a)\), it puts
the old \(a\) including all its outgoing arcs into \(clone\). (This
amounts to making a copy of the heap cell.)
In practice, the efficiency of this approach is considerably improved
by observing that the cloning need only be done during a very short
phase of the collection cycle. The most complex aspect of this
approach is the computation of the pointers into the heap that come
from the local fields (stack, registers) of the Mutator; it has to be
ensured that during this phase no simultaneously changed pointers get
lost. In \cite{Azatchi+2003,LevPet:2001} this is performed by a
technique called ``snooping'': essentially all pointers that are
changed during this phase are treated as roots. (We will come back to
this in \Sec{sec:Roots}.)

For the technical details of this approach we refer to
\cite{Azatchi+2003}. Suffice it to say that almost all of our
subsequent refinements -- which we start from the specification
\(MicroStep\) -- could also descend from \(Snapshot\).  Technically,
we could combine \(Snapshot\) with our various refinements of
\(MicroStep\) by forming a pushout.

\emph{Remark}: By showing that such an effectively working
implementation exists, we have implicitly shown that the specification
\(Snapshot\) is consistent with the specification \(MicroStep\); that
is, the refinement is admissible.





\section{Garbage Collection in Dynamic Graphs}
\label{sec:Refinements}
\label{sec:Dynamic-graphs}

We now take specific properties of garbage collection into account --
but still on the ``semi-abstract'' level of sets and graphs.

First we note that our specification of garbage collection using sets
and set inclusion is a trivial instance of the lattice-oriented
specification in the previous section. Therefore all results carry
over to the concrete problem. The morphism is essentially defined by
the following correspondences:

\begin{center}
$\Phi = %
\bigg[\begin{tabular}{l@{\quad$\mapsto$\quad}l}
\(A\)  & \(Set(Node)\)\\
\(=<\)  & \(\subseteq\)\\
\(f_{i}(s)\) & \(f(G_{i})(s) = s \cup G_{i}.sucs(s) = s \cup \bigcup_{a\in s}G_{i}.sucs(a)\)\\
\(r\) & \(G_{0}.roots\)
\end{tabular}\bigg]
$%
\end{center}

\begin{itemize}
\item The basis now is a sequence of graphs \(G_{0}, G_{1}, G_{2}, ...\) which are due to the activities of the Mutator.
\item The function \(f(G_{i})(s) = s \cup \bigcup_{a\in s}G_{i}.sucs(a)\) adds to the set \(s\) all its direct successors. (We will retain the shorthand notation \(f_{i} = f(G_{i})\) in the following.)
\end{itemize}

\begin{figure}[htbp]
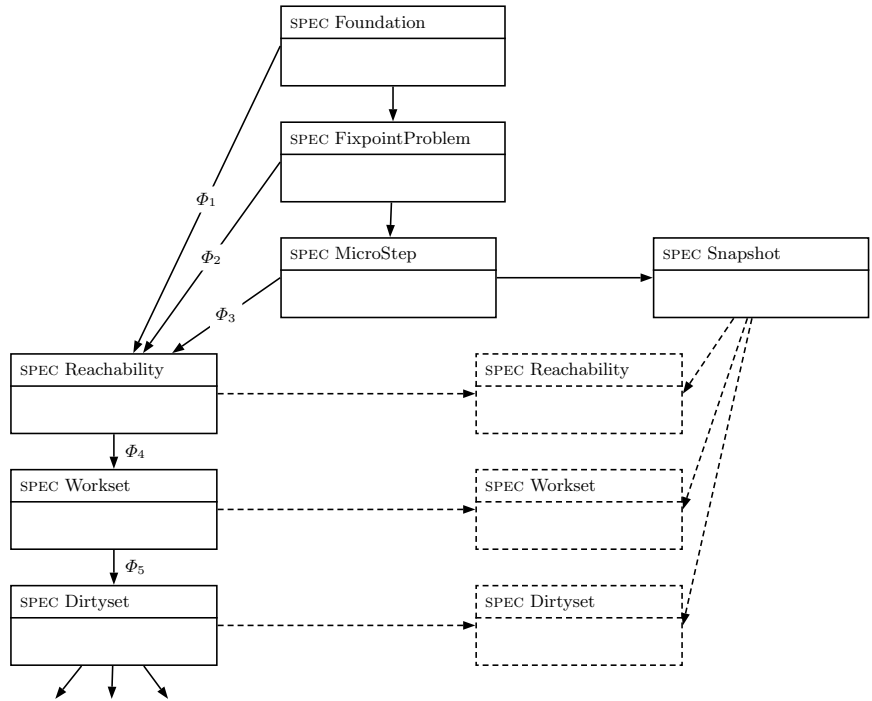
%
\begin{center}%
\Unit{0.01\linewidth}%
\psscalebox{.75}{%
\begin{tabular}{l@{\hspace{1cm}}l@{\hspace{2.5cm}}l}%
& \Spec Foundation[25,0]={Foundation}{} &
\\[.5cm]
& \Spec Fixpoint[25,0]={FixpointProblem}{} &
\\[.5cm]
& \Spec Micro[24,0]={MicroStep}{} & {\Spec Snap[24,0]={Snapshot}{}}
\\[.5cm]
\Spec Reach[23,0]={Reachability}{} & \mc2c{\DSpec Reach2[23,0]={Reachability}{}}
\\[.5cm]
\Spec Workset[23,0]={Workset}{} &  \mc2c{\DSpec Workset2[23,0]={Workset}{}} 
\\[.5cm]
\Spec Dirtyset[23,0]={Dirtyset}{} & \mc2c{\DSpec Dirtyset2[23,0]={Dirtyset}{}}
\\[.2cm]
\pnode(8mm,0){A1}\pnode(18mm,0){A2}\pnode(28mm,0){A3}
\end{tabular}%
\ncline{->}{Foundation}{Fixpoint}%
\ncline{->}{Fixpoint}{Micro}%
\ncline{->}{Reach}{Workset}\Aput{$\Phi_{4}$}%
\ncline{->}{Workset}{Dirtyset}\Aput{$\Phi_{5}$}%
\ncline{->}{\Left{Foundation}}{{Reach}}\mput*{$\Phi_{1}$}%
\ncline{->}{\Left{Fixpoint}}{{Reach}}\mput*{$\Phi_{2}$}%
\ncline{->}{\Left{Micro}}{Reach}\mput*{$\Phi_{3}$}%
\ncline{->}{\Right{Micro}}{Snap}
\psset{linestyle=dashed}
\ncline{->}{{Snap}}{\Right{Reach2}}
\ncline{->}{{Snap}}{\Right{Workset2}}
\ncline{->}{{Snap}}{\Right{Dirtyset2}}
\ncline{->}{\Right{Reach}}{\Left{Reach2}}
\ncline{->}{\Right{Workset}}{\Left{Workset2}}
\ncline{->}{\Right{Dirtyset}}{\Left{Dirtyset2}}
\psset{linestyle=solid}
\ncline{->}{Dirtyset}{A1}
\ncline{->}{Dirtyset}{A2}
\ncline{->}{Dirtyset}{A3}
}
\caption{Roadmap of refinements}
\label{fig:Concrete-refinement}
\end{center}%
\end{figure}

\Fig{fig:Concrete-refinement} illustrates the road map through our
essential refinements. The upper half shows the refinements that have
been performed in the previous \Sec{sec:Fixpoints} on the abstract
mathematical level of lattices and fixed points. The lower half shows
the refinements on the semi-abstract level of graphs and sets that
will be presented in this section. Finally, the right side of the
diagram points out that all further developments could also be
combined (by way of pushouts) with the sidetrack of the snapshot
approach of \Sec{sec:Snapshots}.

\medskip

\begin{lemma}[Morphism abstract \(->\) concrete]\label{lemma:Morphism}
Under the above morphism \(\Phi\) all axioms of the abstract specifications \(Foundation\), \(FixpointProblem\) and \(MicroStep\) hold for the more concrete specifications of graphs and sets (see \Fig{fig:Concrete-refinement}).
\end{lemma}

\emph{Proof}: We show the three morphism properties $\Phi_{1}, \Phi_{2}, \Phi_{3}$ in turn.

$\Phi_{1}$: The proof is trivial, since the monotonicity axiom \Mon is a direct consequence of the definition of \(\Phi(f_{i})\). Axiom \Fp is just a definition.

$\Phi_{2}$: To foster intuition, we first consider the special case \(x=r\): the morphism translates:

\begin{spec}
\ \Caged\ \stackrel{\Phi}{|->}\ \[\%
    f(G_{i+1})\big(\fh(G_{i})(r)\big) \subseteq \fh(G_{i})(r) 
        <=>   \- $\fh(G\sb{i})(r) = live\sb{i}$, def. of $\Phi(f\sb{i})$
    \big(live_{i} \cup \bigcup_{a\in live_{i}}G_{i+1}.sucs(a)\big) \subseteq live_{i}
        <=>   \- $(A\sb{1}\cup...\cup A\sb{n}) \subseteq B \Leftrightarrow \forall i: A\sb{i}\subseteq B$
    \forall a\in live_{i}: G_{i+1}.sucs(a) \subseteq live_{i}
\end{spec} 

In order to prove this last property, i.e. \(\forall a\in live_{i}: G_{i+1}.sucs(a) \subseteq live_{i}\), we must consider all nodes \(a\in live_{i}\) and all (sequences of) actions that the Mutator can use to effect the transition \(G_{i} \leadsto G_{i+1}\). We distinguish the two possibilities for \(a\in live_{i}\):

(1) \(a\in G_{i}.freelist\): Then there are two subcases (which base
on the reasonable constraint that nodes in the freelist and newly
created nodes do not have ``wild'' outgoing pointers):

\begin{spec}[plain]
(1a)\ a \in G_{i+1}.freelist, then G_{i+1}.sucs(a)\subseteq G_{i+1}.freelist \%
                                         \subseteq G_{i}.freelist \subseteq live_{i}
(1b)\ a \in G_{i+1}.active (caused by addNew), then \[ G_{i+1}.sucs(a)=\0; \ now (2)\ applies \]
\end{spec}

(2) \(a\in G_{i}.active\): Then there are three subcases for \(b\in G_{i+1}.sucs(a)\): 

\begin{spec}
(2a)\ (a->b) \in G_{i}.arcs |- b\in G_{i}.active \subseteq  live_{i} 
(2b)\ (a->b)\ created by addArc(a,b) |- b\in G_{i}.active \subseteq  live_{i}
(2c)\ (a->b)\ created by addNew(a) |- b\in G_{i}.freelist \subseteq  live_{i}
\end{spec}

If we start this line of reasoning not from the roots \(r\) but from a
superset \(x \supseteq r\), then we need to consider supersets
\(\Lh_{i} \supseteq live_{i}\) (where the hat shall indicate that
these sets are closed under reachability) and prove \(\forall a\in
\Lh_{i}: G_{i+1}.sucs(a) \subseteq \Lh_{i}\). Evidently the reasoning
in (1) and (2) applies here as well. But now there is a third case:

(3) \(a\in G_{i}.dead\). In this case there is no operation of the
Mutator that could change the successors of \(a\) (since all
operations require \(a\in active\)). Hence \(G_{i+1}.sucs(a) =
G_{i}.sucs(a)\). Due to the closure property we have \(a\in\Lh_{i} =>
G_{i}.sucs(a) \subseteq \Lh_{i}\). The above equality then entails
also \(G_{i+1}.sucs(a) \subseteq \Lh_{i}\).

$\Phi_{3}$: The morphism $\Phi$ translates the axioms \Start and \Step
into

\quad\(s_{i} \subseteq s_{i} \cup \bigcup_{a\in s_{i}}G_{i}.sucs(a)\)

This is trivially fulfilled such that the constraint on the choice of
\(s_{i+1}\) is well-defined.  \\

\Eop

When considering the last specification \emph{Micro-Step} in
\Fig{fig:Microstep-Spec} then we have basically shown that any
sequence \(s_{0}, s_{1}, s_{2}, ...\) that fulfills the constraints
\Start and \Step solves our task. But we have not yet given a
\emph{constructive} algorithm for building such a sequence. In the
next refinement steps \(\Phi_{4}\) and \(\Phi_{5}\) we will proceed
further towards such a constructive implementation (actually to a
whole collection of implementation variants) by adding more and more
constraints to our specification. Each of these refinements
constitutes a design decision that narrows down the set of remaining
implementations.

\subsection{Worksets (``Wavefront'')}
\label{sec:Worksets}

As a first step towards more constructive descriptions we return to
the standard idea of ``worksets'' (sometimes also referred to as
``wavefront''), which has already been illustrated Program
\ref{alg:FP1}, and in the examples in \Sec{sec:Example and Bug}. This
refinement is given in \Fig{fig:Workset}.

\begin{spec}[box=\progA,width=.85\textwidth]
EXTEND MicroStep
b_{0}, b_{1}, b_{2}, ... : A	      \- completely treated (``black'')
w_{0}, w_{1}, w_{2}, ... : A	      \- partially treated (``workset'' or ``gray'')

s_{i} = (b_{i} \uplus w_{i})			\- partitioning into black and gray
\fh_{i}(s_{i}) = b_{i} \cup \fh_{i}(w_{i}) \qquad\qquad\& \WS \- additional constraint
THM w_{n} = \0 => \fh_{n}(s_{n}) = b_{n}  \qquad\&\Term\- termination condition
\end{spec}%
\begin{figure}[htbp]
\begin{center}%
\Unit{0.01\linewidth}
\Spec InitialSpec[0,25]={Workset}{\usebox\progA}
\caption{The workset approach}
\label{fig:Workset}
\end{center}
\end{figure}

The partitioning \(s_{i} = (b_{i} \uplus w_{i})\) arises naturally
from the definition of the workset, as in Program \ref{alg:FP1}. But
the additional axiom \WS is a major constraint! It essentially states
that the closure \(\fh_{i}(s_{i})\) of the current approximation
\(s_{i}\) shall be primarily dependent on the closure of the workset
\(w_{i}\). This reduces the design space of the remaining
implementations considerably -- but from a practical viewpoint this is
no problem, since we only exclude inefficient solutions.

The theorem \Term stated in the specification provides a termination
condition for the later implementations that is far more efficient
than our original termination criterion \(f_{n}(s_{n}) =s_{n}\).

\emph{An important observation}: It is easily seen that the subtle
error situation illustrated in \Fig{fig:Intuition-Problem} in
\Sec{sec:Example and Bug} violates the axiom \WS. Therefore any
further refinement of the specification \(Workset\) cannot exhibit
this error. In other words: If we derive all our implementations as
offsprings of the specification \(Workset\) in \Fig{fig:Workset}, then
we are certain that the bug cannot occur!

\emph{A major problem}: Unfortunately, just introducing sufficient
constraints for excluding error situations is not enough. Consider the
situation of \Fig{fig:Intuition-Problem} in \Sec{sec:Example and Bug}.
We have to ensure that the Mutator cannot perform the two operations
\(addArc(A,E)\) and \(delArc(D,E)\) without somehow keeping the axiom
\WS intact. This necessitates for the first time that the Mutator
cooperates with the Collector, thus introducing constraints for the
Mutator. (Even though these constraints may be hidden in the component
\(Store\), they do have an implicit influence on the Mutator's
working.)%

As has already been pointed out in \Sec{sec:Example and Bug}, there
are three principal possibilities to resolve this problem:

\begin{itemize}
\item One can stop the Mutator until the Collector has finished (Section \ref{sec:Classical-Fixpoint}).
\item One can put \(A\) or \(E\) into the workset, when \(addArc(A,E)\) is executed.
\item One can put \(E\) into the workset, when \(delArc(D,E)\) is executed.
\end{itemize}

Each of these ``solutions'' keeps the axiom \WS intact, but they have
problems. Stopping the Mutator is unacceptable, since this destroys
the very idea of having Mutator and Collector work concurrently. In
both of the other cases the Mutator adds elements to the workset,
while the Collector is taking them out of the workset.  Naive
implementations of this specification would not guarantee termination.

In the following we will present a number of refinements for solving
this problem. These refinements are the high-level formal counterparts
of solutions that can be found in the literature and in realistic
production systems for the JVM and .Net.

\subsection{``Dirty Nodes''}
\label{sec:Dirty-nodes}

One can alleviate the stop times for the Mutator by splitting the
workset into two sets, one being the original workset of the
Collector, the other assembling the critical nodes from the Mutator.
This is shown in \Fig{fig:Dirtyset}.  The new axiom \Dirty is similar
to \WS using the partitioning \(w_{i} = (g_{i} \uplus d_{i})\).

\begin{spec}[box=\progA,width=.90\textwidth]
EXTEND Workset
g_{0}, g_{1}, g_{2}, ... : A	   \- partially treated by Collector (``gray'')
d_{0}, d_{1}, d_{2}, ... : A	   \- introduced by Mutator (``dirty'')

s_{i} = (b_{i} \uplus g_{i} \uplus d_{i})		\- partitioning into black, gray and dirty
\fh_{i}(s_{i}) = b_{i} \cup \fh_{i}(g_{i}) \cup \fh_{i}(d_{i}) \quad\& \Dirty \- closure condition
THM g_{n} = \0 => \fh_{n}(s_{n}) = b_{n} \cup \fh_{n}(d_{n})  \quad\&\DirtyTerm\- intermediate termination condition
\end{spec}%
\begin{figure}[htbp]
\begin{center}%
\Unit{0.01\linewidth}
\Spec InitialSpec[0,25]={Dirtyset}{\usebox\progA}
\caption{Introducing ``dirty'' nodes}
\label{fig:Dirtyset}
\end{center}
\end{figure}

This specification can be implemented by a Collector that successively
treats the gray nodes in \(g_{i}\) until this set becomes empty (which
can be guaranteed). But -- by contrast to the earlier algorithms --
this does not yet mean that all live nodes have been found. As the
theorem \DirtyTerm shows we still have to compute \(\fh_{i}(d_{i})\).
But this additional calculation tends to be short in practice, and the
Mutator can be stopped during its execution.  Consequently,
correctness has been retained and termination has been ensured.

The Mutator now adds ``critical'' nodes to the ``dirty'' set
\(d_{i}\). In order to keep the set \(d_{i}\) as small as possible one
does not add all potentially critical nodes to it: as follows from
axiom \Dirty, black or gray nodes need not be put into \(d_{i}\). And
since \(d_{i}\) is a set, nodes need not be put into it repeatedly.
Actually, when the Mutator executes \(addArc(a,b)\) with \(a\notin
s_{i}\) (``\(a\) is still before the wavefront''), then axiom \Dirty
would allows us the choice of putting \(a\) into \(d_{i}\) or not
(similarly for \(b\).  Commonly, \(a\) is simply added to \(d_{i}\).

\subsection{Implementing the Step $s_{i} \mapsto s_{i+1}$}
\label{sec:Step-implementation}

So far all our specifications only impose the constraint \Step (see
\(MicroStep\) in \Fig{fig:Microstep-Spec}) on their implementations,
that is:

\strut\qquad\(s_{i} < s_{i+1} =< f_{i}(s_{i})\quad \/ \quad s_{i} =
f_{i}(s_{i})\)

The actual computation of the step \(s_{i} |-> s_{i+1}\) has to be
implemented by some function \(step\). For this function we can have
different degrees of granularity:

\begin{itemize}
\item In a \emph{coarse-grained} implementation we pick some node
\(x\) from the gray workset and add all its non-black successors to
the workset. Then we color \(x\) black.

This variant is simpler to implement and verify, but it entails a long
atomic operation. The corresponding write barrier slows down the
standard working of the Mutators.

\item In a \emph{fine-grained} implementation we treat the individual
pointer fields within the current (gray) node \(x\) one-by-one. In our
abstract setting this means that we work with the individual arcs.

This makes the write barrier shorter and thus increases concurrency,
but the implementation and its correctness proof become more
intricate.
\end{itemize}

\begin{figure}[htbp]
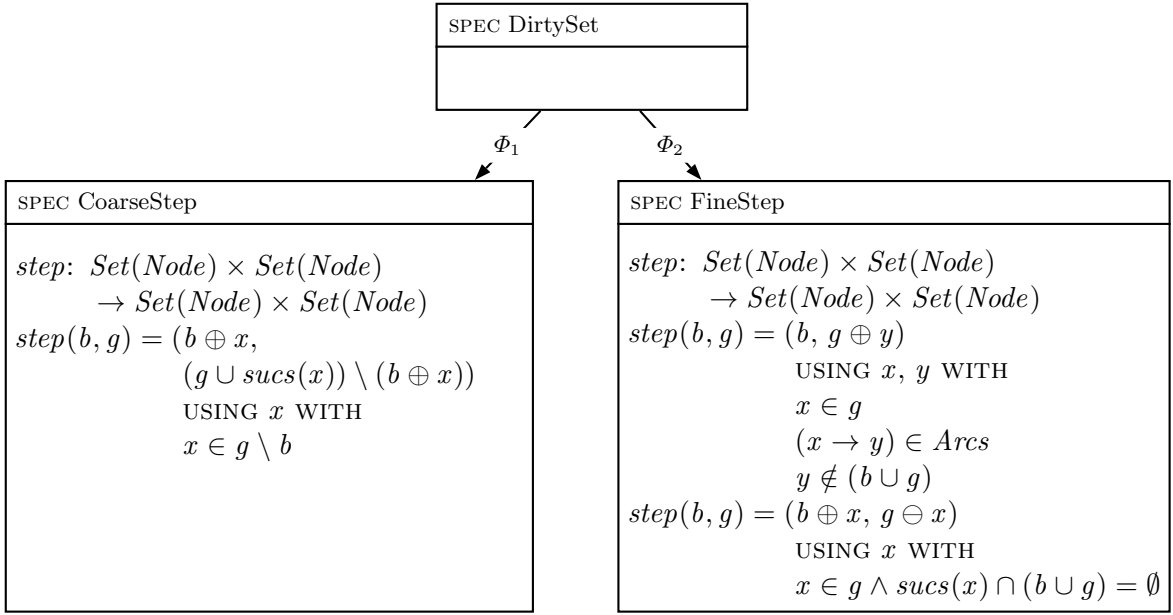
%
\begin{center}%
\Unit{0.01\textwidth}%
\begin{tabular}{c@{\hspace{1cm}}c}%
\mc{2}{c}{\Spec Dirty[26,0]={DirtySet}{}}
\\[.8cm]
\Spec Coarse[44,36]={CoarseStep}{\ \usebox\progA} & 
\Spec Fine[46,36]={FineStep}{\ \usebox\progB}
\end{tabular}%
\ncline{->}{Dirty}{Coarse}\mput*{$\Phi_{1}$}
\ncline{->}{Dirty}{Fine}\mput*{$\Phi_{2}$}
\caption{Step functions of different granularities}
\label{fig:Step-refinements}
\end{center}%
\end{figure}

On our abstract level we treat this design choice by way of two
different refinements. This is depicted in \Fig{fig:Step-refinements}
(where the shorthand notation \(... USING x WITH p(x)\) entails that
the property only has to hold when such an \(x\) exists).

\begin{spec}[width=.43\textwidth,box=\progA]
step: \[Set(Node) ** Set(Node) 
        -> Set(Node) ** Set(Node)\]
step(b,g) = (\[b\oplus x, 
				(g\cup sucs(x)) \setminus (b\oplus x))
               USING x WITH
               x\in g\setminus b
             \]

\hbox{\ }

\hbox{\ }

\hbox{\ }

\end{spec}%
\begin{spec}[width=.45\textwidth,box=\progB]
step: \[Set(Node) ** Set(Node) 
        -> Set(Node) ** Set(Node)\]
step(b,g) = (\[b, g\oplus y)
               USING x, y WITH
               x\in g
               (x -> y) \in Arcs 
               y \notin (b\cup g)\]
step(b,g) = (\[b\oplus x, g\ominus x)
               USING x WITH
               x\in g /\ sucs(x) \cap (b \cup g) =\0
\end{spec}%

\emph{A note of caution.} %
If we apply the morphism \(\Phi\) introduced at the beginning of
\Sec{sec:Refinements} directly, the strict inclusion \(s_{i} <
s_{i+1}\) of axiom \Step would not be provable. Therefore we must
interpret

\strut \qquad \((b,g) < (b^{\prime},g^{\prime})\quad
\stackrel{\Phi}{\mapsto} \quad b \subset b^{\prime} \vee (b=b^{\prime}
\wedge g \subset g^{\prime})\).

\bigskip

But there are still further implementation decisions to be made. Both
\(CoarseStep\) and \(FineStep\) specify (at least partly) how the
\(step\) operation deals with the selected gray node. But this still
leaves one important design decision open: \emph{How are the gray
  nodes selected?}  In the literature we find several
approaches to this task:

\begin{enumerate}
\item \emph{Iterated scanning.} One may proceed as in the original
  paper by Dijkstra et al.~\cite{Dijkstra+1978} and repeatedly scan
  the heap, while applying \(step\) to all gray nodes that are
  encountered. This has the advantage of not needing any additional
  space, but it may lead to many scans over the whole heap, in the
  worst case \(\BigO(N^{2})\) times, and is not considered practical.
	
\item Alternatively one performs the classical recursive graph
  traversal, which may equivalently be realized by an iteration with a
  workset managed as a stack. This allows all the well-known
  variations, ranging from a stack for depth-first traversal to a
  queue for breadth-first traversal. In any case the time cost is in
  the order \(\BigO(|live|)\), since only the live nodes need to be
  scanned. However, there also is a worst-case need for
  \(\BigO(|live|)\) space -- and space is a scarce resource in the
  context of garbage collection.

\item One may compromise between the two extremes and approximate the
  workset by a data structure of bounded size (called a \emph{cache}
  in \cite{Doligez+1994,Doligez+1993}). When this cache overflows one
  has to sacrifice further scan rounds.

\item When there are multiple mutators for efficiency it is necessary
  to have local worksets working concurrently.
\end{enumerate}

\begin{figure}[htbp]
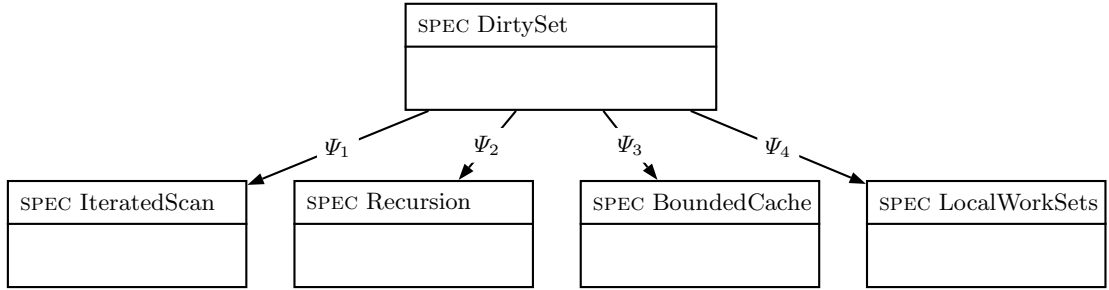
%
\begin{center}%
\Unit{0.01\textwidth}%
\begin{tabular}{c@{\hspace{.5cm}}c@{\hspace{.5cm}}c@{\hspace{.5cm}}c}%
\mc{4}{c}{\Spec Dirty[26,0]={DirtySet}{}}
\\[.8cm]
\Spec Scan[20,0]={IteratedScan}{} & 
\Spec Recursion[20,0]={Recursion}{} &
\Spec Cache[20,0]={BoundedCache}{} &
\Spec Concurrent[20,0]={LocalWorkSets}{} 
\end{tabular}%
\ncline{->}{Dirty}{Scan}\mput*{$\Psi_{1}$}
\ncline{->}{Dirty}{Recursion}\mput*{$\Psi_{2}$}
\ncline{->}{Dirty}{Cache}\mput*{$\Psi_{3}$}
\ncline{->}{Dirty}{Concurrent}\mput*{$\Psi_{4}$}
\caption{Design choices for finding the gray nodes}
\label{fig:Scan}
\end{center}%
\end{figure}

These design choices are illustrated in \Fig{fig:Scan}. (But we
refrain from coding all the technical details.)

It should be emphasized that the refinements \(\Psi_{1}\),
\(\Psi_{2}\), \(\Psi_{3}\), \(\Psi_{4}\) of \Fig{fig:Scan} are
independent of the refinements \(\Phi_{1}\), \(\Phi_{2}\) of
\Fig{fig:Step-refinements}.  This means that we can combine them in
any way we like. The combination of some \(\Phi_{i}\) with some
\(\Psi_{j}\) is formally achieved by a pushout construction as already
mentioned in \Sec{sec:Problems}. In a system like Specware
{\cite{Specware03}} such pushouts are performed automatically.

\medskip

It should be noted that axiom \DirtyTerm in \Fig{fig:Dirtyset}
requires at least one scan in order to perform the cleanup
\(\fh_{n}(d_{n})\) of the dirty nodes after the main marking phase is
completed.

The Collector may also compute (parts of) \(\fh_{i}(d_{i})\) at any
given point in time concurrently with the Mutators. This may make the
finally remaining dirty set \(d_{n}\) smaller and thus speed up the
cleanup operation, which shortens the necessary pauses for the
Mutators. These interim computations are harmless as long as the
termination of the Collector's main marking phase does not depend on
the set \(d_{i}\) becoming empty.

\subsection{Heap Partitioning (Generations, Cards, Pages etc.)}
\label{sec:Heap-partitioning}

The necessary scanning of the ``dirty nodes'' described in
\Sec{sec:Step-implementation} above motivates a refinement that
actually has a whole variety of different applications.  In other
words, the following abstract refinement is the parent of a number of
further refinements that aim at solving different kinds of problems.

In the following we briefly sketch, how such alternative refinements
fit into our abstract framework. In \Secs{sec:Generational-collectors}
and~\ref{sec:Dirty} we will discuss some concrete applications of this
paradigm, in particular
\begin{itemize}
\item \emph{generational} garbage collectors;
\item \emph{dirty cards};
\item \emph{dirty pages}.
\end{itemize}

We integrate these techniques into our approach by means of
\emph{superimposing} a further structuring on the graph. This has
already been hinted at in
\Figs{fig:Intuition-Start}--\ref{fig:Intuition-Problem} in
\Sec{sec:Example and Bug}, where the planes are further partitioned
into areas.

\medskip

\begin{definition}[Graph partitioning]\label{def:Graph-partitioning}
A \emph{partitioning} of the graph is given by splitting the set of nodes into disjoint subsets:

\qquad\strut\(Nodes = N_{1} \uplus ... \uplus N_{k}\)

The subsets \(N_{i}\) are called \emph{cards} (following the
terminology of e.g.~\cite{PriDet:2000}).
\end{definition}

These cards can be used to optimize the computation of the dirty nodes
without compromising the correctness of the algorithm. To this end we
need to introduce the constraint shown in the specification
\(DirtyCards\) in \Fig{fig:Dirtycard}.

\begin{spec}[box=\progA,width=.85\textwidth]
EXTEND Dirtyset
\TYPE\ Card = Set(Node)
N_{1}, ..., N_{k} : Card	   \- cards
Nodes = N_{1} \uplus ... \uplus N_{k}		\- cards partition the node set

dirty: Card -> Bool				\- ``dirty'' property of cards
d_{i} \subseteq \bigcup\, { N_{j}:\ dirty(N_{j})\ } \qquad\qquad \& \DirtyCards \- constraining dirty nodes
\end{spec}%
\begin{figure}[htbp]
\begin{center}%
\Unit{0.01\linewidth}
\Spec InitialSpec[0,24.5]={DirtyCards}{\usebox\progA}
\caption{Introducing ``dirty'' cards}
\label{fig:Dirtycard}
\end{center}
\end{figure}

The axiom \DirtyCards establishes the constraint that the dirty nodes can only lie on dirty cards. (This constraint has to be obeyed by the Mutator.)

The axiom \DirtyTerm in \Fig{fig:Dirtyset} requires the cleanup computation of \(\fh_{n}(d_{n})\) after the main marking phase has been completed. For this cleanup the new axiom \DirtyCards entails a considerable speedup, since only a subset of the cards \(N_{j}\) need to be scanned.





\section{Dynamic Root Sets}
\label{sec:Roots}

In the previous sections we have performed the transition from
classical fixed points in static graphs (relating to stop-the-world
collectors) to dynamic fixed points in changing graphs (relating to
concurrent collectors). However, we still utilize the inherent
assumption that the Collector starts from a \emph{fixed root set}
\(r\). Alas, this assumption -- which is also contained in the
original papers by Dijkstra, Steele and others
\cite{Dijkstra+1978,Steele:1975} -- can not be maintained in practice
due to the following reasons (see
e.g.~\cite{Doligez+1994,Doligez+1993,Azatchi+2003}):

\begin{enumerate}
\item The local data of the Mutator (registers and stack) are indeed
\emph{local}; that is, the Collector has \emph{no access} to them!

\item The synchronization between the Collector and the Mutators
requires \emph{write barriers}. The corresponding overhead may be
tolerable for heap accesses, but it is certainly out of the question
for local stack or register operations.
\end{enumerate}

As a consequence of the first observation the Mutators have to
participate actively in the garbage collection process, at least
during the start phase of each collection cycle. And the second
observation rules out certain solutions due to their unacceptable
overhead. So the challenge is to maximize concurrency in the presence
of these constraints.

In the following we will show how these issues fit into our overall
framework.

\subsection{Modeling Local Data}

In accordance with our earlier levels of abstraction we devise the
following modeling for the Mutator-local data: all local data of a
Mutator (registers and stack) are considered as one large node, which
we call the pre-root \(\rho_{m}\). The potentially quite large set
\(sucs(\rho_{m})\) then represents all pointers out of the local stack
and registers into the heap.%
\footnote{%
  Here it pays that we model pointers more abstractly as a (multi)set
  of arcs. This is much more concise and elegant than speaking about
  ``objects'' and their ``slots'' for pointers, as it is usually done
  in the literature.%
} %

From now on let us assume that there are \(q\) Mutators \(M_{1}, ...,
M_{q}\) with pre-roots \(\rho_{1}, ..., \rho_{q}\). The global
variables are represented by the pre-root \(\rho_{0}\); they are
accessible by the Collector.

\medskip 

\begin{definition}[Pre-roots; local roots]\label{def:Pre-roots}
  Each Mutator possesses a \emph{pre-root} \(\rho_{m}\). The
  successors of this pre-root are referred to as the \emph{Mutator's
    local roots}: \(r_{m} = sucs(\rho_{m})\).

  Moreover, the \emph{global variables} (which are accessible to the
  Collector) are represented by the pre-root \(\rho_{0}\) and the
  corresponding roots by \(r_{0} = sucs(\rho_{0})\).

  The set of all pre-roots is denoted as \(\rho = {\rho_{0}, \rho_{1},
    ..., \rho_{q}}\). The set of all roots is defined as \(r =
  sucs(\rho) = r_{0} \cup \bigcup_{m\in Mut}r_{m}\).
\end{definition}

The set \(r\) introduced in \Def{def:Pre-roots} above essentially
plays the role of the start value \(s_{0}\) in the specification
\(MicroStep\) of \Fig{fig:Microstep-Spec}. Hence, we might rephrase
the central property \AimN of this specification (after applying the
morphism \(\Phi\) from \Sec{sec:Refinements}):

\qquad\strut \(\exists n: \fh_{n}(r) \subseteq s_{n} \subseteq
\fh_{0}(r) \quad WHERE r = sucs(\rho) = sucs(\rho_{0}) \cup
\bigcup_{m\in Mut}sucs(\rho_{m})\).

However, due to the subtle difficulties that are caused by the
concurrent activities of Collector and Mutators we should retreat to a
more fundamental rephrasing of our overall task.

\subsection{Computation of the Roots}
\label{sec:Root-computation}

The abstract modeling introduced in \Def{def:Pre-roots} at the
beginning of this section allows us to retain all the other modeling
aspects of the preceding sections. In particular the concept of
varying graphs \(G_{0}, G_{1}, G_{2}, ...\) and the thus induced
functions \(f_{0}, f_{1}, f_{2}, ...\) can be applied to the
computation of the root set \(r\) without changes.

However, for reasons that will become clear in a moment, we need to
make one change. Since the local registers and stacks of the Mutators
are not part of the heap, they must not undergo the mark and sweep or
the copying process. In our mathematical modeling we therefore no
longer consider the reflexive-transitive closure \(\fh\) but only the
non-reflexive transitive closure, which we denote as \ft. As a
consequence, the function \(f\) should not be inflationary either.
Hence, the morphism \(\Phi\) at the beginning of
\Sec{sec:Dynamic-graphs} now sets \(f_{i}(s) = \bigcup_{a\in
  s}G_{i}.sucs(a)\). Consequently, the axiom \Step in the
specification \(MicroStep\) has to be changed to

\qquad\strut \(s_{i} < s_{i+1} =< s_{i} \sqcup f_{i}(s_{i}) \/ f_{i}(s_{i}) =< s_{i}\)\qquad \Step

With these changes (for which all previous proofs work unchanged
except for minor notational adaptations) we can now reformulate the
garbage collection task slightly differently (see
\Fig{fig:Rephrased-task}, where the numbers are the same as in the
original specifications).

\begin{spec}[box=\progA,width=.85\textwidth]
\ft(x) = LEAST s:\ f(x) =< s /\ s = f(s) \qquad\& \Fp      \- transitive closure 

s_{0}, s_{1}, s_{2}, ... : A	      \- sequence of approximations

s_{i} < s_{i+1} =< s_{i} \sqcup f_{i}(s_{i}) \/ f_{i}(s_{i}) =< s_{i}\qquad\& \Step \- computation step

THM\ \exists\, n:\ \ft_{n}(\rho) =< s_{n} =< \ft_{0}(\rho) \qquad\& \AimN  \- $\mathit{live}\sb{n} \subseteq s\sb{n} \subseteq \mathit{live}\sb{0}$
\end{spec}%
\begin{figure}[htbp]
\begin{center}%
\Unit{0.01\linewidth}
\Spec InitialSpec[0,21]={Micro-Step$^{\prime}$}{\usebox\progA}
\caption{The ``micro-step approach''}
\label{fig:Rephrased-task}
\end{center}
\end{figure}

A major difference between this specification \(MicroStep^{\prime}\)
and the original specification \(MicroStep\) is the omission of the
axiom \Start, which determines the start value \(s_{0}\). This start
value is no longer a constant, but now has to be computed from the
pre-roots.  This computation is slightly intricate, since the graph is
undergoing continuous changes. To be more precise, we have the
following scenario:

\begin{itemize}
\item There are \(q\) mutators \(M_{1}, ..., M_{q}\).

\item When Mutator \(M_{i}\) computes its local roots \(r_{i}\) from
  its pre-root \(\rho_{i}\), then the graph is in some stage
  \(G_{j}\).

\item During the root computation the mutator stops its other
  activities. And the other mutators cannot access the mutator's local
  data. Hence the outgoing arcs from the local data into the heap
  remain unchanged throughout the local root computation. Hence we
  obtain \(r_{i} = f_{i}(\rho_{i}) = G_{j}.sucs(\rho_{i})\).%
  \footnote{%
    Note that this does \emph{not} mean that the graph remains
    invariant throughout the computation of the local root set
    \(r_{i}\). On the contrary, the other mutators will usually change
    the graph continuously. But these changes do not affect the
    specific set \(G_{j}.sucs(\rho_{i})\).  } %
\end{itemize}

Based on these observations, the set \(s_{0} = r\) that is computed by
the mutators essentially is

\qquad\strut\(r = f_{0}(\rho_{0}) \cup f_{1}(\rho_{1}) ... \cup f_{q}(\rho_{q})\) \qquad // too naive

However, this naive approach doesn't work all the time as we will show in the following.

\begin{figure}[htbp]
\begin{center}%
\Unit{1mm}
\def\M#1{%
\begin{pspicture}(0,0)(20,08)
\psframe(0,0)(20,08)
\rput(10,04){$M_{#1}$}
\end{pspicture}%
}
\newcommand\Nde[2][white]{%
\begin{pspicture}(0,0)(12,06)
\psframe[fillstyle=solid,fillcolor=#1](0,0)(12,06)
\rput(06,03){$#2$}
\end{pspicture}%
}
\begin{tabular}{c@{\hspace{2cm}}c}
\begin{pspicture}(0,0)(60,30)

\rput[lt](00,30){\rnode{M1}{\M{1}}}
\rput[lt](40,30){\rnode{M2}{\M{2}}}

\rput[lb](26,10){\rnode{A}{\Nde[lightgray]{a}}}
\rput[l](39,13){$\in r_{1}$}

\rput[lb](06,00){\rnode{B}{\Nde{b}}}

\ncline{->}{M1}{A}
\ncline{->}{A}{B}
\end{pspicture}
& 
\begin{pspicture}(0,0)(60,30)

\rput[lt](00,30){\rnode{M1}{\M{1}}}
\rput[lt](40,30){\rnode{M2}{\M{2}}}

\rput[lb](26,10){\rnode{A}{\Nde[lightgray]{a}}}
\rput[l](39,13){$\in r_{1}$}

\rput[lb](06,00){\rnode{B}{\Nde{b}}}

\ncline{->}{M1}{A}
\ncline[linestyle=dashed]{->}{M1}{B}\Bput[0pt]{\(addArc(\rho_{1},b)\)}

\pnode(24,08){point}
\nccurve[linestyle=dotted,angleA=290,angleB=320]{M2}{point}\Aput[0pt]{\(delArc(a,b)\)}
\end{pspicture}
\end{tabular}
\caption{A potential error}
\label{fig:Potential-error}
\end{center}
\end{figure}
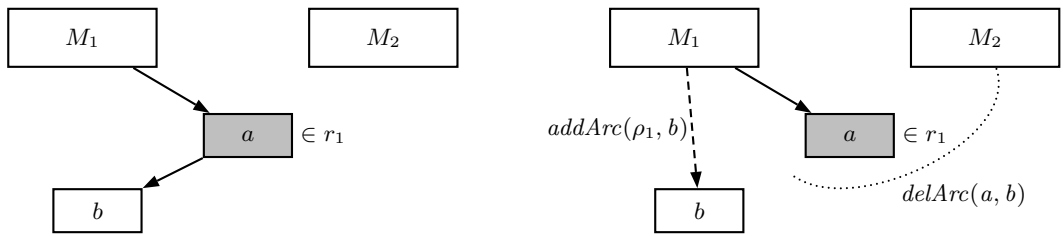

\paragraph{A problem.}%
Looking at our central correctness property \AimN we would wish that
the equality \(\ft(\rho) = \fh(r)\) holds. Alas, this is not
necessarily the case. To see this, consider the example of
\Fig{fig:Potential-error} (adapted from \cite{Doligez+1994}).

Suppose \(M_{1}\) has computed its (only) local root \(a\),
i.e.~\(r_{1} = {a}\) (left side of \Fig{fig:Potential-error}). Then it
resumes its normal activities, which happen to load a pointer to \(b\)
into a local variable or register; this is modeled as
\(addArc(\rho_{1},b)\) on the right side of \Fig{fig:Potential-error}.
When the Collector now starts its recycling activities, we have
\(b\notin r_{1}\), even though \(b\in sucs(\rho_{1})\). So the
Collector starts from a wrong root set.

It is easily seen that this can indeed be disastrous: suppose that
before the start of the Collector some Mutator \(M_{2}\) (or \(M_{1}\)
itself) deletes the pointer from \(a\) to \(b\) (right side of
\Fig{fig:Potential-error}). Then \(b\) will indeed be considered
garbage, even though it is still reachable from \(M_{1}\).

This is the same problem as the one that we had already encountered in
\Fig{fig:Intuition-Problem} of \Sec{sec:Example and Bug}. However,
there is a difference: the problematic pointers now are not caused by
heap operations but by operations on the local variables and
registers. For reasons of efficiency we do not want to slow down these
local operations by wrapping them into read or write barriers.%
\footnote{%
  In some systems there is not even enough information to distinguish
  pointers from other values such that an abundance of operations
  would have to be engulfed into this protective overhead of barriers.
} %
This would be particularly unpleasant, since the protection is only
needed during an extremely short phase (namely the root marking),
while the overhead would be hindering permanently.

In \cite{Doligez+1994} further problems are illustrated that could
occur, when the Mutator is e.g.~interrupted between the
\(addArc(\rho_{1},b)\) and the \(delArc(a,b)\) operation for such a
long time that the Collector performs a whole collection cycle.

\paragraph{Towards a solution.}%
We enforce the invariant

\qquad\strut\(\ft_{i}(\rho) \subseteq \fh_{i}(s_{i})\).

Then when the computation terminates with the fixpoint \(s_{n} =
f_{n}(s_{n})\),  the fixpoint properties immediately entail

\qquad\strut\(\ft_{n}(\rho) \subseteq s_{n}\).

This can be rephrased as

\qquad\strut\(live_{n} \subseteq s_{n}\).

which is the main correctness criterion for the collector (as was
stated in property \AimN of \Fig{fig:Microstep-Spec}).

When looking at the critical situation of \Fig{fig:Potential-error},
we can see that there are three perceivable solutions for establishing
this requirement:

\begin{enumerate}
\item We can stop all other activities of the mutators during the
  local-root computation.

  This is very safe, but it clearly has the disadvantage of causing
  long pauses. In \cite{Doligez+1994} it is shown how this solution
  can be achieved using three handshakes that ensure that the
  Collector and all Mutators are in sync.

\item We could request that the operation \(addArc(\rho_{1},b)\) puts
  \(b\) into the set \(s_{i}\).

  This is a correct solution, but it has the disadvantage that we need
  a \emph{read barrier} for loading heap pointers into local variables
  or registers. Even though this read barrier will be a single if-test
  during most of the time, it does add overhead.

\item We could request that the operation \(delArc(a,b)\) puts \(b\)
  into the set \(s_{i}\).

  This is a correct solution, which needs a \emph{write barrier} on
  heap operations. This is not so bad, since this write barrier
  already exists for handling the other problems encountered in the
  previous sections. Still, there are disadvantages: marking nodes at
  the very moment of their deletion will create floating garbage in
  the majority of cases. Moreover, the operation \(delArc(a,b)\) has
  to touch the node \(a\), since it changes one of its slots; but the
  marking additionally has to touch another node, namely \(b\). This
  adds to the overhead.
\end{enumerate}

We should mention that a fourth kind of solution is possible in the
``snapshot-at-the-beginning'' approaches (see \Sec{sec:Snapshots}).
This is demonstrated in the so-called ``sliding-views'' approach of
\cite{Azatchi+2003}.

\begin{enumerate}\setcounter{enumi}{3}
\item The operation \(delArc(a,b)\) produces a clone of the node \(a\).

  The advantage here is that less floating garbage is generated and
  that only the object \(a\) needs to be touched by the write barrier.
  But one has the overhead of the storing and managing the snapshot.
\end{enumerate}





\section{Real-world Considerations}
\label{sec:Real-world}

It is well known that realistic garbage collectors -- in particular
concurrent or parallel ones -- exhibit a huge amount of technical
details that are ultimately responsible for the size and complexity of
the verification efforts. The pertinent issues cover a wide range of
questions such as:

\begin{itemize}
\item What are the exact read and write barriers?
\item How do we treat the references in the global variables, the
stacks and the registers?
\item Where do we put the marker bits (in mark-and-sweep collectors)
or the forward pointers (in copying collectors)?
\end{itemize}

Evidently we do not have the space here to address these questions in
detail. But we should at least indicate the path towards their
solution within our method. Therefore we list here some of the
technical features contained in realistic product-level collectors and
show how they fit into our framework.

\subsection{The Mutators' Capabilities}
\label{sec:Mutator-capabilities}

In \Sec{sec:Architecture} we have limited the Mutators' behavior to
the three operations \(addArc\), \(delArc\) and \(addNew\). By
contrast, Doligez et al.~\cite{Doligez+1994,Doligez+1993} use eight
operations, some of which are modified in other approaches
\cite{Domani+2000}. We may group their operations as follows:

\begin{itemize}
\item \emph{move, load}: local data transfers (stack, registers) and
read access to heap cells;
\item \emph{reserve, create}: obtain space from freelist; create cell
in that space;
\item \emph{fill, update}: write into a new/existing cell;
\item \emph{cooperate, mark}: synchronize with Collector; mark local
roots.
\end{itemize}

This design breaks the usual \emph{allocate} operation into the three
separate operations \emph{reserve, create, fill}. Moreover it treats
\emph{fill} differently from \emph{update}, since a new object is
guaranteed to be still unknown to other Mutators and therefore can be
handled with less synchronization overhead. (But in the approach of
\cite{Domani+2000} this distinction is abolished.) This diversity and
granularity is important for concrete discussions about issues such as
Mutator-local mini heaps and other implementation details. But for our
modeling approach and its refinement and correctness considerations
our three basic operations cover the essential aspects.
Technically speaking, the eight operations of Doligez et al.~can be
obtained by refining our low-level models even further.

Moreover, Doligez et al.~\cite{Doligez+1994,Doligez+1993} and many of
their successor papers use a notation like \texttt{heap[x,i]} to refer
to the $i$-th slot in the object $x$ in the heap. Our model achieves
the same effect with a more mathematical attitude by considering
individual arcs \((a->b)\). In other words, \((a->b_{1})\), \dots,
\((a->b_{k})\) model the slots of the object \(a\). Our notation
allows a more flexible treatment of the question of whether objects
are uniform or can have varying numbers of successor slots.

\subsection{Availability of a Runtime System}
\label{sec:Runtime-system}

\Sec{sec:Architecture} introduces an architecture with a component
\(Store\) that represents the memory management used in modern runtime
systems such as .Net or JVM, based on old ideas from the realm of
functional languages such as ML or Haskell. Such a component provides
only indirect access to the heap such that it is relatively easy to
integrate read or write barriers, handshakes and other organizational
means. The vast majority of the newer papers therefore target these
kinds of architectures.

In ``uncooperative languages'' such as C or \Cpp things are more
intricate and it is not surprising that only a few papers address
their demands, e.g.~\cite{Boehm+1991}. The difficulties caused by such
uncooperative languages are overcome by using the capabilities of the
underlying operating system such as the page table to introduce
concepts like \emph{dirty pages}. Moreover, one has to deal with lots
of floating garbage, since -- due to the lack of typing information --
many non-pointer values have to be treated conservatively as if they
were pointers. Last but not least there are also more intricate
synchronization issues between the Collector and the
Mutators. Benchmarks indicate that the overhead in such uncooperative
languages is considerably higher than that in cooperative languages.

\subsection{Generational Collectors}
\label{sec:Generational-collectors}

In \Sec{sec:Heap-partitioning} we have shown that we can superimpose
the graph with an additional structuring such that the set of nodes is
partitioned into subsets: \(Nodes = N_{1}\uplus ... \uplus
N_{k}\). Such a partitioning is fully compatible with our correctness
considerations and the pertinent refinements:

\emph{Generational} garbage collectors partition the nodes according
to their ``age'', where the age usually reflects the number of
collection cycles that the node has survived.

Whereas in traditional garbage collectors the nodes are physically
moved to another area, this is no longer possible in concurrent
collectors due to the large overhead that the pointer tracking and
synchronization would require. Therefore one usually only defines the
generations logically. A non-moving solution for concurrent collectors
is presented by Domani et al.~\cite{Domani+2000a} basing on earlier
work of Demers et al.~\cite{Demers+1990}.

Printezis and Detlefs \cite{PriDet:2000} describe a concurrent
generational collector that has been implemented as part of the SUN
Research JVM.
The different generations often use different collectors. For the
young generation, where nodes tend to die fast, copying collectors are
the technique of choice, whereas in older generations a mark-and-sweep
approach works better \cite{PriDet:2000}.

The most critical issue in generational collectors is the treatment of
old-to-young references, since the whole point of generations is NOT
to touch the elder generations in most collection cycles. To cope with
these references (that are caused by the Mutators' activities) one
usually employs the dirty-cards or dirty-pages techniques that we
discuss in the section.

\subsection{``Dirty Areas'' (Cards, Pages, \dots)}
\label{sec:Dirty}

As has been pointed out in \Sec{sec:Heap-partitioning} the
disturbances caused by the Mutators can be encapsulated into a concept
of ``dirty areas'', which allows a compromise between accuracy and
efficiency.

\begin{itemize}
\item \emph{Cards} are often used to speed up the scanning by
constraining it to memory areas that actually may need scanning. This
is done by using a \emph{dirty bit} for each card, on which a mutator
has performed some update: when the dirty bit is set, the card needs
to be scanned.

Cards and dirty bits are used (possibly under a different name) in
connection with generational garbage collectors \cite{PriDet:2000} but
also to cope with problems caused by multiple mutators. They can be
based on very efficient write barriers. For example, the SUN
ResearchVM \cite{PriDet:2000} uses the two-instruction write barrier
proposed in \cite{Hoe:1993}.

\item \emph{Pages} taken from the virtual-page mechanism of the
underlying OS together with \emph{dirty bits} can be used for
uncooperative languages like C and \Cpp, where no runtime system
exists that nicely separates the application programs from the memory
management \cite{Boehm+1991}. However, as is reported in
\cite{HosMos:1993}, the use of cards is more efficient than the use of
page-protection-based barriers.
\end{itemize}

\subsection{Privacy of Local Data; Local Heaps}
\label{sec:Local-data}

One of the primary goals of many concurrent-collector designs is to
keep the stop times of the mutators to a minimum (in practice within
an order of magnitude of 2ms; see e.g. \cite{Azatchi+2003}). Here the
greatest barrier is the global synchronization, when the mutators
derive the local roots from their pre-roots (local registers and
stack). \emph{The longest time that a thread waits for garbage
collection is the time for it to mark the objects directly reachable
from its stack} \cite{Domani+2000}.

Based on the observation that the major percentage of heap dynamics
comes from short-lived and small objects, the global synchronization
can be made less frequent by providing every mutator with its own
local ``mini heap'' \cite{Doligez+1994,HawPet:2009}. Then the mutator
only needs to interact with the global heap in order to acquire a new
mini heap, when the old one is full, or in order to store large
objects. Otherwise it uses a very simple allocation scheme in its
local heap, e.g. using a so-called ``bumper-pointer'' technique
\cite{HawPet:2009}. This design, which is used e.g. in the IBM JVM,
also has the advantage of being cache-friendly
\cite{Barabash+2005,Borman:2002,KerPet:2006}.

\subsection{Detailed Memory Management}
\label{sec:Detailed-memory-management}

There is a plethora of little details to be considered in real-world
garbage collectors, of which the following list gives but a small
selection.

\begin{itemize}
\item \emph{Object size.} The original garbage collectors by McCarthy
  \cite{McCarthy:1960} or Dijkstra \cite{Dijkstra+1978} are based on
  the assumption of fixed-size heap cells. But in reality these cells
  come in all sizes and internal layouts. There all kinds of solutions
  for this problems such as using freelists of different sizes,
  splitting large objects into smaller pieces, using Mutator-local
  heaps for small objects and so forth. The size information is either
  stored with the object or inferred from the object's class. And so
  forth.

\item{Coalescing.} The sweeping phase should try to combine
consecutive free junks into one large free junk in order to alleviate
the fragmentation problem. This is easy in non-concurrent collectors,
but is more complex in concurrent collectors, since now the Collector
and the Mutators compete for the same resource, namely for the removal
of cells from the \(freelist\) \cite{PriDet:2000}.

\item \emph{Marker and pointer management.} All the algorithms use
various kinds of markers -- for example the colors black, gray, white
or the dirty bits -- and various pointers -- for example the forward
pointers in copying collectors or the clone pointers in the
``sliding-views'' approaches. A typical technique for handling such
problems is e.g. described in \cite{HawPet:2009}, where the vtable
pointer, which is the first word of every object, is overwritten for
the forward pointer needed in the copying collector. These pointers
can be distinguished, since they point into disjoint (and known)
storage areas.

\item \emph{Sets, markers and bitmaps.} Many markers actually
represent the membership of the node in a certain set. This can either
be implemented by a marker bit in each object or by a global set
representation using a bitmap.

\item \emph{Coping with ``no-information''.} Most garbage collection
approaches nowadays address the JVM or DotNet, where all memory
accesses of application programs are indirect, since they are handled
by a memory manager. Therefore all the garbage collection activity can
be bundled in the runtime system. Old systems based on C or \Cpp do not
exhibit this luxury. There one may at best use work-arounds such as
employing the virtual-paging mechanism of the underlying OS by making
pages ``dirty'', whenever they are written to
\cite{Boehm+1991}. Another problem here is that many integer values
have to treated as if they were pointers, this way producing a lot of
floating garbage.
\end{itemize}





\section{Conclusion}
\label{sec:Conclusion}

We have shown how the main design concepts in contemporary concurrent
collectors can be derived from a common formal specification.  The
algorithmic basis of the concurrent collectors required the
development of some novel generalizations of classical fixpoint
iteration theory.  We hope to find a wide variety of applications for
the generalized theory, as there has been for the classical theory.
This is of interest since the reuse of abstract design knowledge
across application domains is a key factor in the economics of formal
derivation technology.  Alternative refinements from the basic
algorithm lead to a family tree of concurrent collectors, with shared
ancestors corresponding to shared design knowledge.  While our
presentation style has been pedagogical, the next step is to develop
the derivation tree in a formal derivation system, such as Specware.

\textbf{Acknowledgment.}  We are grateful to Erez Petrank and Chris
Hawblitzel, with whom one of us (pp) enjoyed intensive discussions at
Microsoft Research. Their profound knowledge on the challenges of
practical real-world garbage collectors motivated us to push our
original high-level and abstract treatment further towards concrete
and detailed technical aspects -- although we realize that we may
still be on a very abstract level in the eyes of true practitioners.



\bibliographystyle{plain} \bibliography{references}
\end{document}